\documentclass[preprintnumbers,10pt,a4paper,superscriptaddress,aps,pre, twocolumn,floatfix]{revtex4-2}

\usepackage{geometry}
\usepackage{mathtools}
\usepackage{graphicx}
\usepackage{dcolumn}
\usepackage{bm}
\usepackage{makecell}
\usepackage{xcolor}
\usepackage{enumerate}
\usepackage[caption=false]{subfig}
\usepackage{float}
\usepackage[unicode=true,pdfusetitle,bookmarks=true,bookmarksnumbered=false,bookmarksopen=false,
breaklinks=false,pdfborder={0 0 1},backref=false,colorlinks=true]
{hyperref}
\hypersetup{linkcolor=blue,urlcolor=blue,citecolor=blue}
\usepackage{physics}
\usepackage{geometry}

\newcommand{\be}{\begin{equation}}
	\newcommand{\ee}{\end{equation}}
\newcommand{\bd}{\begin{displaymath}}
	\newcommand{\ed}{\end{displaymath}}
\newcommand{\BE}{\begin{eqnarray}}
	\newcommand{\EE}{\end{eqnarray}}

\newcommand{\boldpsi}{{\mbox{\boldmath $\psi$}}}

\newcommand{\avg}[1]{\left\langle{#1}\right\rangle}

\begin{document} 
	
	
	\title{Generalised correlations in disordered dynamical systems: Insights from the many-species Lotka--Volterra model}
	
	\author{Sebastian Castedo}
	\email{sebastian.castedo@phys.ens.fr}
	\affiliation{Department of Physics and Astronomy, School of Natural Sciences, The University of Manchester, Manchester M13 9PL, UK}
	
	\author{Joshua Holmes}
	\email{josh.holmes2112@gmail.com}
	\affiliation{Department of Physics and Astronomy, School of Natural Sciences, The University of Manchester, Manchester M13 9PL, UK}
	
	\author{Joseph W. Baron}
	\email{joseph-william.baron@phys.ens.fr}
	\affiliation{Laboratoire de Physique de l’Ecole Normale Sup\'erieure, ENS, Universit\'e PSL,
		CNRS, Sorbonne Universit\'e, Universit\'e de Paris, F-75005 Paris, France}
	
	\author{Tobias Galla}
	\email{tobias.galla@ifisc.uib-csic.es}
	\affiliation{Instituto de F\'isica Interdisciplinar y Sistemas Complejos, IFISC (CSIC-UIB), Campus Universitat Illes Balears, E-07122 Palma de Mallorca, Spain}

	\date{\today}
	
	\begin{abstract}
		In the study of disordered systems, one often chooses a matrix of independent identically distributed interaction coefficients to represent the quenched random couplings between components, perhaps with some symmetry constraint or correlations between diagonally opposite pairs of elements. However, a more general set of couplings, which still preserves the statistical interchangeability of the components, could involve correlations between interaction coefficients sharing only a single row or column index. These correlations have been shown to arise naturally in systems such as the generalised Lotka-Volterra equations (gLVEs). In this work, we perform a dynamic mean-field analysis to understand how single-index correlations affect the dynamics and stability of disordered systems, taking the gLVEs as our example. We show that in-row correlations raise the level of noise in the mean field process, even when the overall variance of the interaction coefficients is held constant. We also see that correlations between transpose pairs of rows and columns can either enhance or suppress feedback effects, depending on the sign of the correlation coefficient. In the context of the gLVEs, in-row and transpose row/column correlations thus affect both the species survival rate and the stability of ecological equilibria. 
	\end{abstract}
	
	\maketitle
	
	\section{Introduction}
	The theory of spin glasses was first developed 50 years ago to describe disordered magnetic materials. Routes taken to studying these systems included both equilibrium and non-equilibrium approaches \cite{mezard1987, altieri2024introduction, altieri2020dynamical}, such as the famous replica approach and dynamic mean-field theory. The use of tools from disordered systems then rapidly expanded beyond condensed matter physics \cite{charbonneau2023spin}, and to date includes neural networks \cite{rajan2006eigenvalue, kuczala2016eigenvalue, aljadeff2015transition, sompolinsky1988chaos}, optimisation and machine learning \cite{ros2019complex, biroli2023generative}, game theory and the modelling of financial markets \cite{coolen_kuehn_sollich, Coolen_MG}.   
	
	One of the areas in which ideas from the theory of disordered systems has enjoyed a recent surge of attention is theoretical community ecology. In the tradition of May's seminal work \cite{may, maybook}, one can try to understand which kinds of \textit{statistics} of species interactions permit a stable coexistence. In recent years, the effect of various kinds of interaction structures on stability (and also community composition) has been investigated, using dynamic mean-field theory (DMFT) approaches \cite{bunin2016interaction, galla2018dynamically,roy2019numerical}, static replica approaches \cite{birolibunin}, and results from random matrix theory \cite{allesinatang1, allesinatang2}. For example, the roles of model aspects such as hierarchy \cite{poley1}, complex network structure \cite{poley2, park, aguirre}, physical space \cite{garcia2024interactions, baron2020dispersal}, temporal effects \cite{pigani2022delay, suweis2023generalized} and demographic noise \cite{altieri2021properties}, amongst others, have been considered. 
	
	Some of this work focuses on the analysis of generalised Lotka-Volterra equations (gLVEs). This is a popular model of many-species ecological communities, which has been shown to replicate the time evolution of real communities quite well (see for example Ref. \cite{nguyen2024inferring} and works therein). In the context of the gLVEs, it has been shown that the interactions between coexisting species can exhibit intricate statistical features, owing to the constraints that coexistence imposes \cite{poley2, barbier2021fingerprints}. In particular, one finds that correlations emerge in the surviving community (after initial extinctions) between interaction coefficients that share only a single row or column index, even when they are not present in the original community \cite{bunin2016interaction,baron_et_al_prl}.
	
	In the traditional treatment of disordered systems, complications such as these types of correlations are typically ignored, since the phenomena of interest can often be described by models with independent and identically distributed couplings, usually supplemented with a symmetry constraint. For example, the energy level statistics of heavy nuclei are in agreement with the predictions of the Gaussian orthogonal random matrix ensemble \cite{weidenmuller2009random, haq1982fluctuation}, and simple p-spin glass models provide a useful mechanism for ergodicity breaking and glassy dynamics \cite{crisanti1993spherical}. However, in the context of complex ecosystems, one is more concerned with understanding all the various effects that can promote or jeopardise stability. Since novel single-index correlations can arise in the interaction coefficients organically in ecological models, it is therefore natural to ask what their impact is on the dynamics and the ultimate fate of the community when they are not artificially excluded from the initial community interactions.   
	
	In this work, we perform a dynamic mean-field analysis of the generalised Lotka-Volterra equations with novel single-index (in-row, in-column and transpose row-column) correlations between interaction coefficients. In doing so, we show that in-row correlations lead to a greater amount of effective noise in the system, while transpose row-column correlations can lead to either the suppression or amplification of feedback loops, depending on the sign of the correlation coefficient. These observations are general, and apply to any dynamical model with quenched disorder. In the context of the gLVEs, we then show how these effects can lead to a greater rate of extinction of species, the enhancement of stability of the surviving community, and the suppression/reinforcement of runaway abundance growth. 
	
	The remainder of the paper is structured as follows: In Sec.~\ref{sec:model} we introduce the generalised Lotka-Volterra equations with generalised correlations. In Sec.~\ref{sec:DMFT} we describe how dynamic mean-field theory can be used to understand the collective influence of the wider system on a single species, and we extract the effect of the additional single-index correlations. We also analyse the time-independent unique fixed-point solution to the effective single-species process, and determine conditions for the stability of this solution. In Sec.~\ref{sec:results} we investigate the specific influence that the single-index correlations have on the stability of the system, providing also an intuitive explanation for the effects that we observe (Sec.~\ref{sec:intuition}). We then discuss our results and conclude in Sec.~\ref{sec:disc}.

	\section{Model definitions}
	\label{sec:model}

	\subsection{Dynamics}
	We study a model ecosystem consisting of $N$ species, with time-dependent abundances $x_1,\dots,x_N\geq 0$. The abundances obey a system of  generalised Lotka-Volterra equations \cite{bunin2016interaction,bunin2017,galla2018dynamically}
	\begin{equation}
		\frac{dx_i(t)}{dt} = r_i x_i(t) \left(K_i - x_i + \sum_{j=1}^N \alpha_{ij}x_j \right),
		\label{eq:LV}
	\end{equation}
	where $r_i$ is the basic growth rate of species $i$, and $K_i$ is the corresponding carrying capacity in absence of all other species. Throughout this work we set $r_i=1$ and $K_i=1$ for all $i$, similar to \cite{bunin2016interaction,bunin2017,galla2018dynamically,galla_notes}. 
	
	The interaction matrix elements $\{\alpha_{ij}\}$ describe the effect that the presence of species $j$ has on the time-evolution of species $i$. These coefficients are taken to be quenched random variables following \cite{bunin2016interaction,bunin2017,galla2018dynamically}, and are fixed throughout the dynamics. We describe how they are selected below.
	
	\subsection{Interaction Matrix}
	The most basic choice for the interaction matrix elements involves drawing each $\alpha_{ij}$ from a Gaussian distribution with common mean and variance, as was done by May \cite{may}. From this, one finds that an increased variance of the interaction coefficients (or `complexity') is destabilising. Allowing for correlations between only diagonally opposed pairs $\alpha_{ij}$ and $\alpha_{ji}$, one then finds that negative correlations of this type (which connote a preponderance of predator-prey pairs) is stabilising \cite{bunin2016interaction,bunin2017,galla2018dynamically}. Indeed, one can add many model ingredients, in the form of more exotic or intricate interaction statistics, and observe how they modify stability. For example, hierarchically-structured interaction matrices \cite{poley1}, or Hopfield-like interactions have previously been considered \cite{rozas}. 
	
	Here, we do not include any particularly exotic interaction hierarchy or structure. Instead, we return to the null model in which all species are statistically equivalent, in the sense that the statistics of the interaction coefficients $\alpha_{ij}$ do not depend on their index. We simply investigate the most general set of interaction statistics that does not have any \textit{a priori} bias towards any one species. Specifically, we write
	\begin{equation}
		\alpha_{ij} =   \frac{\mu}{N} + z_{ij},
		\label{eq:alpha}
	\end{equation}
	where $\mu$ characterises the mean value of the interaction strength between species, and where the $z_{ij}$ are random variables of mean zero. They are taken to have the following variance and correlations,
	\begin{equation}
		\begin{gathered}
			\text{var}(z_{ij}) = \frac{\sigma^2}{N},
			\\\text{corr}(z_{ij},z_{ji}) = \Gamma,
			\\\text{corr}(z_{ij},z_{ki}) = \frac{\gamma}{N},
			\\\text{corr}(z_{ij},z_{ik}) = \frac{r}{N},
			\\\text{corr}(z_{ji},z_{ki}) = \frac{c}{N}.
		\end{gathered}
		\label{eq:new-corr}
	\end{equation}
	Here, $\text{corr}(a,b)\equiv \left(\langle ab \rangle - \langle a\rangle \langle b \rangle \right)/\sqrt{\text{var}(a)\text{var}(b)}$. The quantities $\mu, \sigma^2, \Gamma, \gamma, r$ and $c$ are the main model parameters, along with the number of species $N$. The scaling of the mean, variance and correlations in Eqs.~(\ref{eq:alpha}) and (\ref{eq:new-corr}) with $N$ guarantees a well-defined thermodynamic limit $N\to \infty$. The mathematical analysis of the model is carried out in this limit.

\begin{figure}
		\centering
		\includegraphics[width = 0.65\linewidth]{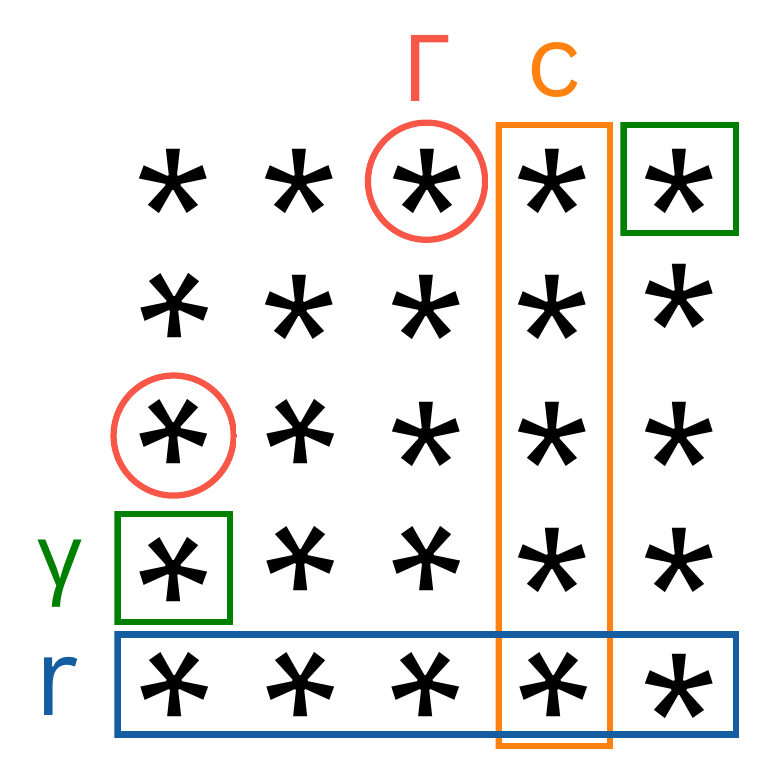}
		\caption{Illustration of the correlations in the interaction matrix. These are correlations within rows (blue, parametrised by $r$), within
			columns (orange, parametrised by $c$), correlations between diagonally opposed entries (red circles, parametrised by $\Gamma$), and transpose row-column correlations (green square, parametrised by $\gamma$). These are described in Eq.~(\ref{eq:new-corr}).}
		\label{fig:new_corr}
	\end{figure}

	The correlations in the interaction matrix are illustrated in Fig.~\ref{fig:new_corr}. The parameter $\Gamma$ describes correlations between diagonally opposed entries $\alpha_{ij}$ and $\alpha_{ji}$. The parameter $r$ captures correlations of elements within the same row in the interaction matrix, and $c$ describes in-column correlations. Finally, $\gamma$ denotes correlations between elements sharing one index, but in opposing positions (e.g. $\alpha_{1,2}$ and $\alpha_{2,3}$). We will refer to these as `transpose row-column' correlations. We do not include correlations between elements that do not share any index. To leading order in $1/N$, these can be considered as identical to adding the same random number to all matrix elements, and can therefore be re-absorbed into the definition of the mean \cite{baron2021eigenvalues}.
	
	\subsection{Numerical Generation of Matrices with Generalised Correlations}
	\label{sec:num}
	
	To construct realisations of the interaction matrix with generalised correlations [Eq.~(\ref{eq:alpha})], we first generate random numbers $\lambda_i, \kappa_i$ ($i=1,\dots,N$), each with mean zero, and with covariance matrix for any pair $\lambda_i,\kappa_i$ as follows,
	\be\label{eq:helpmatrix1}
	\begin{pmatrix}
		\langle \lambda_i^2 \rangle & \langle \lambda_i \kappa_i \rangle\\
		\langle \lambda_i \kappa_i \rangle & \langle \kappa_i^2 \rangle 
	\end{pmatrix}
	= \frac{\sigma^2}{N^2} \begin{pmatrix}
		r & \gamma\\
		\gamma & c 
	\end{pmatrix}.
	\ee
	There are no correlations between the variables $\lambda_i,\kappa_i$ and $\lambda_j,\kappa_j$ for $i\neq j$.
	
	For each pair $i<j$ we also generate Gaussian random variables $w_{ij}$ and $w_{ji}$, each with mean zero, and with covariance matrix
	\be\label{eq:helpmatrix2}
	\begin{pmatrix}
		\langle w_{ij}^2 \rangle & \langle w_{ij} w_{ji} \rangle\\
		\langle w_{ij} w_{ji} \rangle & \langle w_{ij}^2 \rangle 
	\end{pmatrix}
	= \frac{\sigma^2}{N} \begin{pmatrix}
		1-\frac{r+c}{N} & \Gamma-\frac{2\gamma}{N}\\
		\Gamma-\frac{2\gamma}{N} & 1-\frac{r+c}{N} 
	\end{pmatrix}.
	\ee
	There are no correlations between $w_{ij}$ and $w_{k\ell}$ if these two elements are not diagonally opposed to one another. The variables $r$, $c$, $\gamma$ and $\Gamma$ in Eqs.~(\ref{eq:helpmatrix1}) and (\ref{eq:helpmatrix2}) are those in Eq.~(\ref{eq:new-corr}). 
	
	Having generated the $\lambda_i, \kappa_i$, $w_{ij}$ and $w_{ji}$, we set
	\be
	z_{ij} = \lambda_i + \kappa_j+ w_{ij}.
	\ee
	One can directly verify that the $z_{ij}$ generated following this procedure have the properties in Eq.~(\ref{eq:new-corr}). We note that the algorithm can only be used for $r,c\geq 0$, and if $\vert \gamma \vert \leq \sqrt{rc}$.
	
	This method of generating Gaussian random matrices with generalised correlations is also described in the text around Eq.~(S61) in the supplement of Ref. \cite{baron2021eigenvalues}. We note a typographical error in this equation of the earlier reference where a prefactor of $1/N^2$ is given when it should instead be $1/N$.

	\section{Dynamic Mean Field Theory}
	\label{sec:DMFT}
	To derive a dynamical mean-field description for the model, we use the Martin-Siggia-Rose-Janssen-De Dominicis (MSRJD) generating functional formalism \cite{mezard1987,dedominicis1978dynamics,galla_notes}. This broadly consists of setting up a dynamical partition function, which permits one to carry out an average over the disorder. With the introduction of a set of macroscopic order parameters, the partition function takes a saddle-point form and can therefore be evaluated in the limit $N\to\infty$. This procedure is well-established in spin-glass physics \cite{mezard1987, sompolinsky1981dynamic, sompolinsky1982relaxational, crisanti1993spherical}, and has been used on different variants of replicator or Lotka--Volterra dynamics (see for example \cite{galla2018dynamically,sidhomgalla,park}). A detailed tutorial can be found in Ref.~\cite{galla_notes}. Given this, we do not include the full calculation here, but report only the final result. Some further details can be found in Appendix~\ref{sec:full_gen}.  
	
	We also discuss how one can go about deriving the same effective process using the cavity method in Appendix \ref{app:cavity}. We find that the cavity method reproduces one possible interpretation of the effective process that one obtains from the MSRJD approach, but not the other (more easily tractable) one. In this sense, the MSRJD approach proves to be the more transparent and straightforward method in this case, whereas often the two approaches would be comparably convenient. 
	
	\subsection{Effective Process}
	The outcome of the generating-functional analysis is that the aggregate effect of the wider community on a species, which is represented by the interaction term in Eq.~(\ref{eq:LV}), can be encapsulated by the statistical behaviour of a single species. Specifically, one obtains a coloured Gaussian noise term and a memory (or dissipative) term
	\BE\label{eq:eff_proc}
	\dot{x}(t) &=& x(t)\bigg\{ 1-x(t) \nonumber \\
	&&+ \sigma^2 \int_0^t dt' G(t,t')\left[\Gamma x(t') +\gamma M(t')\right]\nonumber \\
	&& + \mu {M}(t) + \eta(t)\bigg\},
	\EE
	along with the following self-consistency relations,
	\BE
	G(t,t') &=& \avg{\frac{\delta x(t)}{\delta \eta(t')}}_*, \nonumber \\
	M(t) &=& \avg{x(t)}_*, \nonumber \\
	\avg{\eta(t)\eta(t')}_* &=& \sigma^2\left[\avg{x(t) x(t')}_* \right. \nonumber \\
	&&\left.~~~+ rM(t)M(t')\right].
	\label{eq:cfs}
	\EE
	Here $\avg{\dots}_*$ denotes an average over the realisations of the effective dynamics in Eq.~(\ref{eq:eff_proc}).
	
	Indeed, one recovers the effective process given in Ref.~\cite{galla2018dynamically} by setting $\gamma = r = 0$ in Eq.~(\ref{eq:eff_proc}). Let us now briefly consider the effect that the inclusion of single-index correlations has on the dynamics of each species. We will discuss the mechanism behind these effects in greater detail in Sections \ref{sec:effM} and \ref{sec:eff_lin_instab}.
	
	Noting that $M(t)\geq 0$ by construction, we see that for $r>0$, the noise correlator in Eq.~(\ref{eq:cfs}) acquires an extra positive term. This means that the effect of the wider community varies to a greater extent between species. Broadly, this results from the fact that (for fixed $\sigma^2$) increasing $r$ (the amount of in-row correlation) increases the uniformity of the interactions $\alpha_{ij}$ of species $i$ with all the other species $j$. Consequently, the variation of the interaction coefficients is greater between rows than within rows, leading to a greater variation in how different species experience the pool of other species.   
	
	We also see that for $\gamma \neq 0$, we obtain an additional memory term in Eq.~(\ref{eq:eff_proc}). The usual memory term proportional to $\Gamma$ encapsulates the fact that changes in an individual species' abundance affect the abundances of other species, and the wider community feeds this disturbance back to the aforementioned species. This term varies from species to species. In contrast, the term proportional to $\gamma$ reflects a more complicated feedback route, which is made possible by the additional interaction correlations. First, a perturbation in the abundance of a species affects the full community of other species. This perturbation then feeds back to the entire community in a more uniform fashion. That is, the term proportional to $\gamma$ is the same for all species. This is discussed in more detail in Appendix~\ref{app:cavity}.
	
	The in-column correlations, parameterised by $c$, are absent from the effective process and self-consistency relations. This is because these correlations do not alter what a particular species experiences. Interaction coefficients that share the same column encapsulate the different effects that a single species has on all the other species in the community. The in-column correlations therefore determine how similarly different species are effected by one particular species. As one might expect, when we consider the effect of the community as a whole on a single species (in a sense, this is the conceptual inverse of what the in-column correlations quantify), the in-column correlations are not relevant. 
	
	\subsection{Fixed Point Analysis}
	The full dynamical solution to the self-consistent effective process in Eqs.~(\ref{eq:eff_proc}) and (\ref{eq:cfs}) can be solved using numerical methods \cite{oppereissfeller,froy}, but analytical progress is generally difficult. In the case of the Lotka--Volterra system, matters simplify in the region of parameter space in which the gLVEs reach a unique stable fixed point (in the limit $N\to\infty$). Extending the analysis described in \cite{bunin2016interaction,bunin2017,galla2018dynamically,galla_notes}, we can determine the statistical features of the fixed point solution and the points in parameter space where the stability of this solution breaks down. 
	
	Fixed points $x^*$ of the effective process are found from the relation
	\BE
	&x^*\Big[1-x^*+ \sigma^2 \chi  (\Gamma x^*+\gamma M^*\big )\nonumber \\
	& +\mu M^* + \eta ^* \Big] = 0.
	\label{eq:fixed_point}
	\EE
	The noise variables $\eta(t)$ become static random variables $\eta^*$ in the fixed-point regime. We have used time-translation invariance, $G(t,t')=G(t-t')$, and we have introduced $\chi=\int_0^\infty d\tau\, G(\tau)$, assuming that this quantity is finite. We also write $M^*$ for the long-time limit of $M(t)$ in the fixed-point phase. The correlation function $C(t,t')$ also becomes constant, and we write $C(t,t')\equiv q$.
	
	For a given value of $\eta^*$, Eq.~(\ref{eq:fixed_point}) admits two solutions: $x^*=0$ and the root of the linear expression inside the square bracket. This latter solution for the fixed-point abundance can only be meaningful when it is non-negative. Similar to previous work on related gLVE systems, we find that this solution is indeed realised whenever it is positive, and that for all other $\eta^*$ the effective species becomes extinct ($x^*=0$). After some algebra (following the lines of \cite{bunin2016interaction, bunin2017,galla2018dynamically,galla_notes}) the self-consistency relations in Eq.~(\ref{eq:cfs}) can be expressed as follows, 
	\begin{equation}
		\begin{split}
			M^* = \sigma \frac{\sqrt{q+r\,(M^*)^2}}{1-\Gamma \sigma^2\chi}&\int_{-\infty}^{\Delta}Dz\,(\Delta - z),\\
			q = \sigma^2 \frac{q+r\,(M^*)^2}{(1-\Gamma \sigma^2\chi)^2}&\int_{-\infty}^{\Delta}Dz\,(\Delta - z)^2,\\
			\chi = \frac{1}{1-\Gamma\sigma^2\chi}&\int_{-\infty}^{\Delta} Dz,
			\label{eqn:op_static}
		\end{split}
	\end{equation}
	where $Dz = dz e^{-z^2/2}/\sqrt{2\pi}$. We have written
	\be
	\Delta =  \frac{1 + M^*[\mu + \gamma \sigma^2\chi]}{\sigma\sqrt{q + r\,(M^*)^2}},\label{eq:delta}
	\ee
	Eqs.~(\ref{eqn:op_static}) and (\ref{eq:delta}) determine $M^*, q, \chi$ and $\Delta$ for given model parameters $\mu,\sigma^2, \Gamma, r, c$ and $\gamma$, and are valid in the phase of stable unique fixed points. An efficient procedure for the numerical solution of the above equations is described in Appendix \ref{sec:sol_proc}. 
	
	For the following discussion, it is useful to define
	\be
	w_{\ell}(\Delta) \equiv \int_{-\infty}^{\Delta}Dz\,(\Delta - z)^{\ell}
	\label{eqn:delta_def}
	\ee
	for $\ell=0,1,2$.
	The quantity
	\be\label{eq:phi}
	\phi=w_0(\Delta)
	\ee
	can then be identified as the fraction of surviving species (see also \cite{bunin2016interaction,bunin2017,galla2018dynamically}).

	\subsection{Stability analysis}
	We now find the conditions under which the assumption of unique stable fixed point breaks down. There are two causes of this: (1) linear instability to perturbation, or (2) the average abundance, $M^*$, can diverge. 
	\subsubsection{Linear instability}\label{sec:lin_instab}
	In order to determine the onset of the linear instability, we performed linear stability analysis along the lines of \cite{opper1992phase,galla2018dynamically}. As in these references, perturbations about extinct species ($x^*=0$) are found to always decay. For solutions of the effective process with a non-zero asymptotic abundance, $x^*> 0$, one finds the following expression for the Fourier transform $\tilde y(\omega)$ of perturbations $y(t)$,
	\begin{equation}
		\langle|\Tilde{y}(0)|^2\rangle = \frac{\phi\left(\sigma^2r{M}^2+1\right)}{[\Gamma \sigma^2 \chi -1]^2 - \phi \sigma^2}.
		\label{eq:fourier}
	\end{equation}
	The left-hand side must be positive by definition, and the right-hand side changes sign when the denominator becomes zero. Hence, the fixed-point solution becomes inconsistent when  $\left(\Gamma \sigma^2 \chi -1\right)^2  = \phi \sigma^2$. The expressions for the order parameters in Appendix~\ref{sec:sol_proc} allow us to simplify this condition further. For given $\mu,\Gamma,c,r,\gamma$, we find that the fixed-point solution is linearly stable only if the variance of the interaction coefficients is sufficiently small, $\sigma^2<\sigma_c^2$, where $\sigma_c^2$ is given by
	\be
	\sigma^2_c(\Gamma, r) = \frac{1}{\phi(\Delta_c)(1+\Gamma)^2}, \label{eq:opper1}
	\ee
	with $\Delta_c$ the solution of
	\be
	\Delta_c = - rw_1(\Delta_c).\label{eq:opper2}
	\ee
	In order to find $\sigma^2_c$, one solves  Eq.~(\ref{eq:opper2}) numerically for $\Delta$, then finds $\phi(\Delta_c)$ via Eq.~(\ref{eq:phi}), which one then inputs into Eq.~(\ref{eq:opper1}). For $r=0$ we find $\Delta_c=0$ and hence $\phi(\Delta_c)=1/2$, and from this, $\sigma_c^2(\Gamma)=2/(1+\Gamma)^2$. This is the familiar onset of linear instability in the gLVE model with only cross-diagonal correlations \cite{opper1992phase,bunin2016interaction,bunin2017,galla2018dynamically}. However, one finds that $\phi(\Delta_c)\neq1/2$ for $r\neq 0$.
	
	\subsubsection{Divergent mean abundance}
	To find the onset of diverging mean abundance, we set $M^*=\infty$ in Eq.~(\ref{eqn:op_static}) [which in turn means we set $1/M^*=0$ in Eq.~(\ref{eq:sol_1_M})]. We then find the following parametric description of the line in the $(\mu,\sigma^2)$-plane at which the mean abundance diverges,
	\begin{subequations}\label{eq:div_M}
		\begin{gather}
			\mu = \frac{\Delta w_2\left(1+r\frac{w_1^2}{w_2}\right)-\gamma\phi w_1}{w_1\left[\phi\Gamma + w_2\left(1+r\frac{w_1^2}{w_2}\right)\right]}, \label{eq:div_M1}\\
			\sigma^2 = \frac{w_2\left(1+r\frac{w_1^2}{w_2}\right)}{\left[\phi\Gamma + w_2\left(1+r\frac{w_1^2}{w_2}\right)\right]^2}. \label{eq:div_M2}
		\end{gather}
	\end{subequations}
	We have suppressed the argument $\Delta$ of the $w_\ell$ on the right-hand sides. Eqs.~(\ref{eq:div_M}) assume fixed values of the remaining model parameters ($\Gamma,c,r,\gamma$), and we use $\Delta$ as a free parameter. The line in the $(\mu,\sigma^2)$-plane at which the divergence occurs is obtained numerically by varying $\Delta$ \cite{poley1, baron_et_al_prl}.
	\section{Implications for stability}
	\label{sec:results}
	
	We now proceed to study the effects of the different types of correlations on the stability and the composition of the ecological community.
	
\begin{figure}[htbp]
		\centering
		\includegraphics[width = 0.9\linewidth]{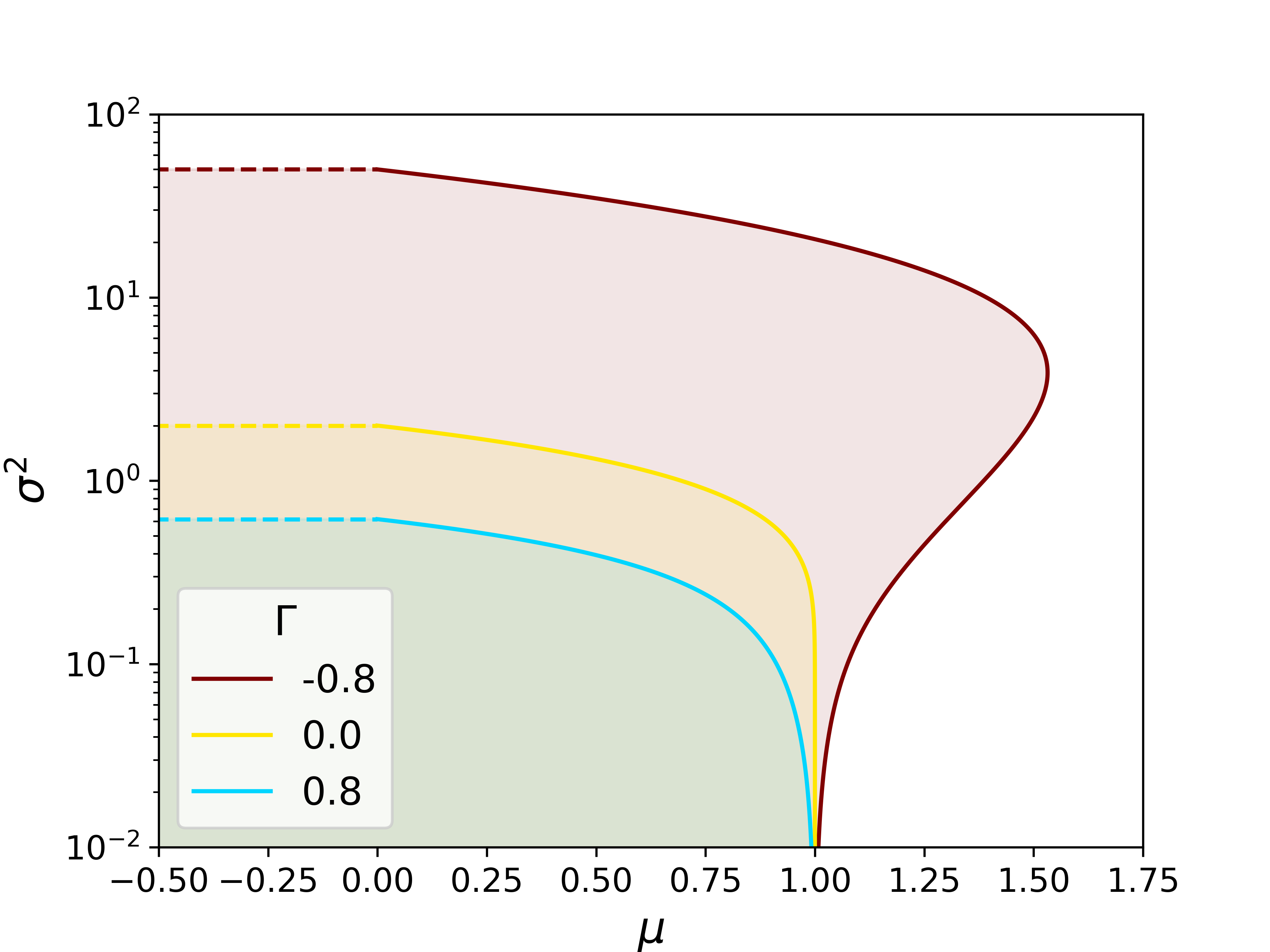}
		\caption{Stability diagram of the gLVE system without generalised correlations in the interaction matrix, as previously reported in \cite{bunin2016interaction,bunin2017,galla2018dynamically,baron_et_al_prl}, for $r=\gamma=0$. The diagram shows the stable region in the space spanned by $\mu$ and $\sigma^2$ for different $\Gamma$. The solid lines correspond to the ${M}\to \infty$ transition and the dashed lines correspond to the onset of linear instability. The stable region is the space to the left of the solid line and below the dashed line. The regions of stability for different values of $\Gamma$ are overlapping.}
		\label{fig:BG}
	\end{figure}

	As a baseline, we briefly describe the known phase diagram of the model with correlations only between diagonally opposed matrix elements \cite{bunin2016interaction,bunin2017,galla2018dynamically}. This is shown in Fig.~\ref{fig:BG} where the shaded areas on the $(\mu,\sigma^2)$ plane indicate the stable phase in which the gLVEs converge to a unique stable fixed point. The different colours in the figure represent different choices of the correlation parameter $\Gamma$. Solid lines indicate the onset of diverging mean abundance, and dashed lines indicate the onset of the linear instability. One particular conclusion that has been drawn from this phase diagram is that negative correlations between diagonally opposed pairs $\alpha_{ij}$ and $\alpha_{ji}$ (i.e., an increased fraction of predator-prey relations) promote stability. 

	We now turn to the effects of the different types of correlations in turn. We first present our observations and then discuss possible mechanistic explanations for these trends in Sections~\ref{sec:effM} and \ref{sec:eff_lin_instab}.
	
	\subsection{In-row correlations (parametrised by $r$)}
\begin{figure}[]
		\centering
		\includegraphics[width=0.9\linewidth]{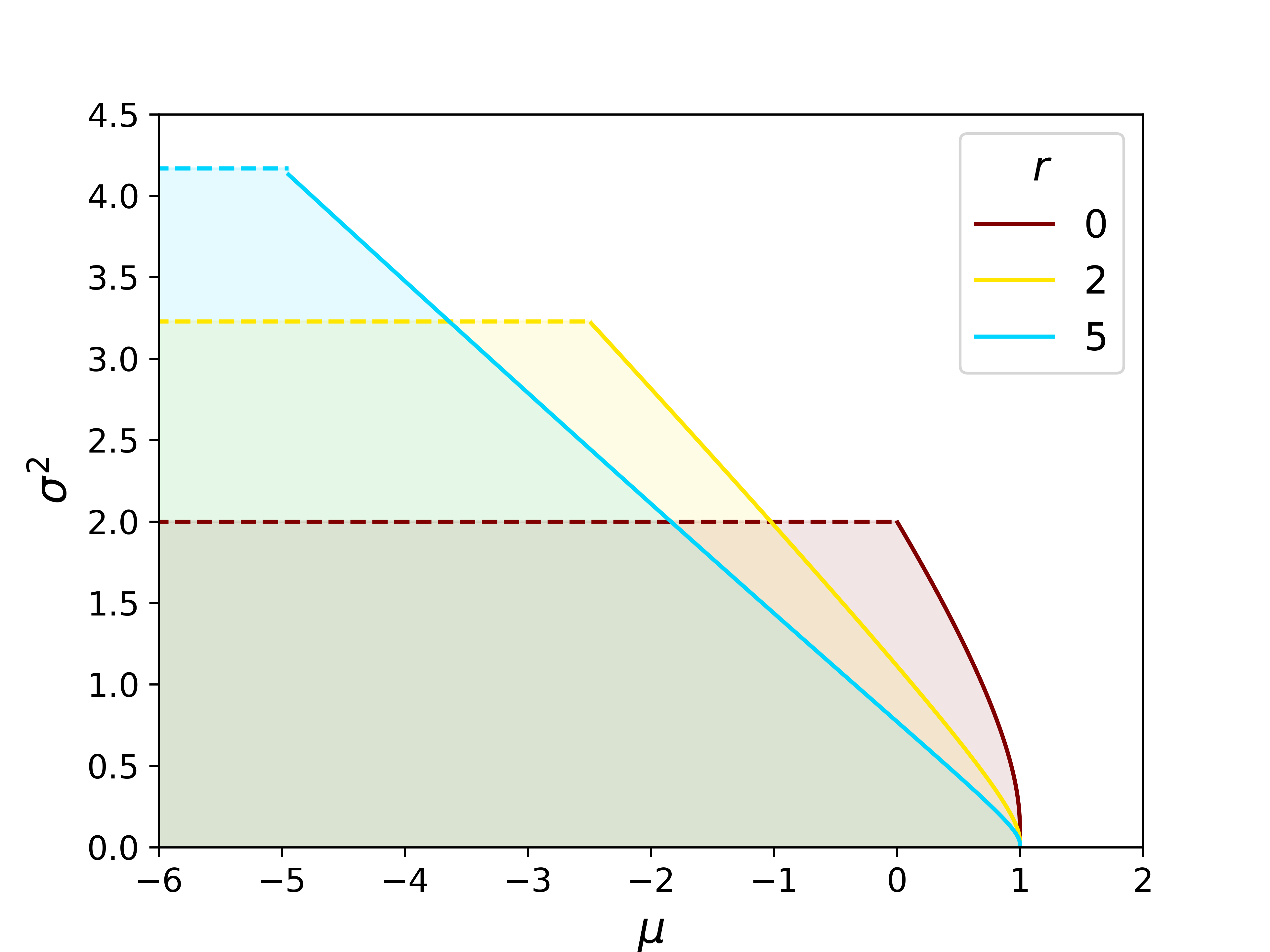}
		\caption{Stability phase diagram for different values of $r$ at fixed $\Gamma=\gamma = 0$. The dashed line indicates the onset of the linear instability, and the solid line shows where the mean abundance first diverges ($M\rightarrow \infty$). The shaded area underneath and to the left of the curves is the stable region. An increase in $r$ leads to a greater degree of linear stability, but promotes the onset of diverging abundances.}
		\label{fig:stab-r}
	\end{figure}

	Fig.~\ref{fig:stab-r} shows the effect that in-row correlations have on the stable region. In particular, increasing in-row correlations moves the line at which abundances diverge to the left in the phase diagram, thus promoting the diverging-abundance transition ($M\to\infty$). At the same time in-row correlations move the linear instability line upwards in Fig.~\ref{fig:stab-r}. This indicates a stabilising effect. We have also verified the effects of in-row correlations on the two types of phase lines in simulations. This is summarised in Sec.~\ref{Sec:num_r} of the Appendix.
	
	In Fig.~\ref{fig:frac-r}, we show the fraction of surviving species, $\phi$, as a function of the in-row correlation parameter $r$. In-row correlations are seen to reduce the number of survivors. We discuss later in Section \ref{sec:eff_lin_instab} how this suppresses the linear instability. 
	
	\begin{figure}
		\centering
		\includegraphics[width=\linewidth]{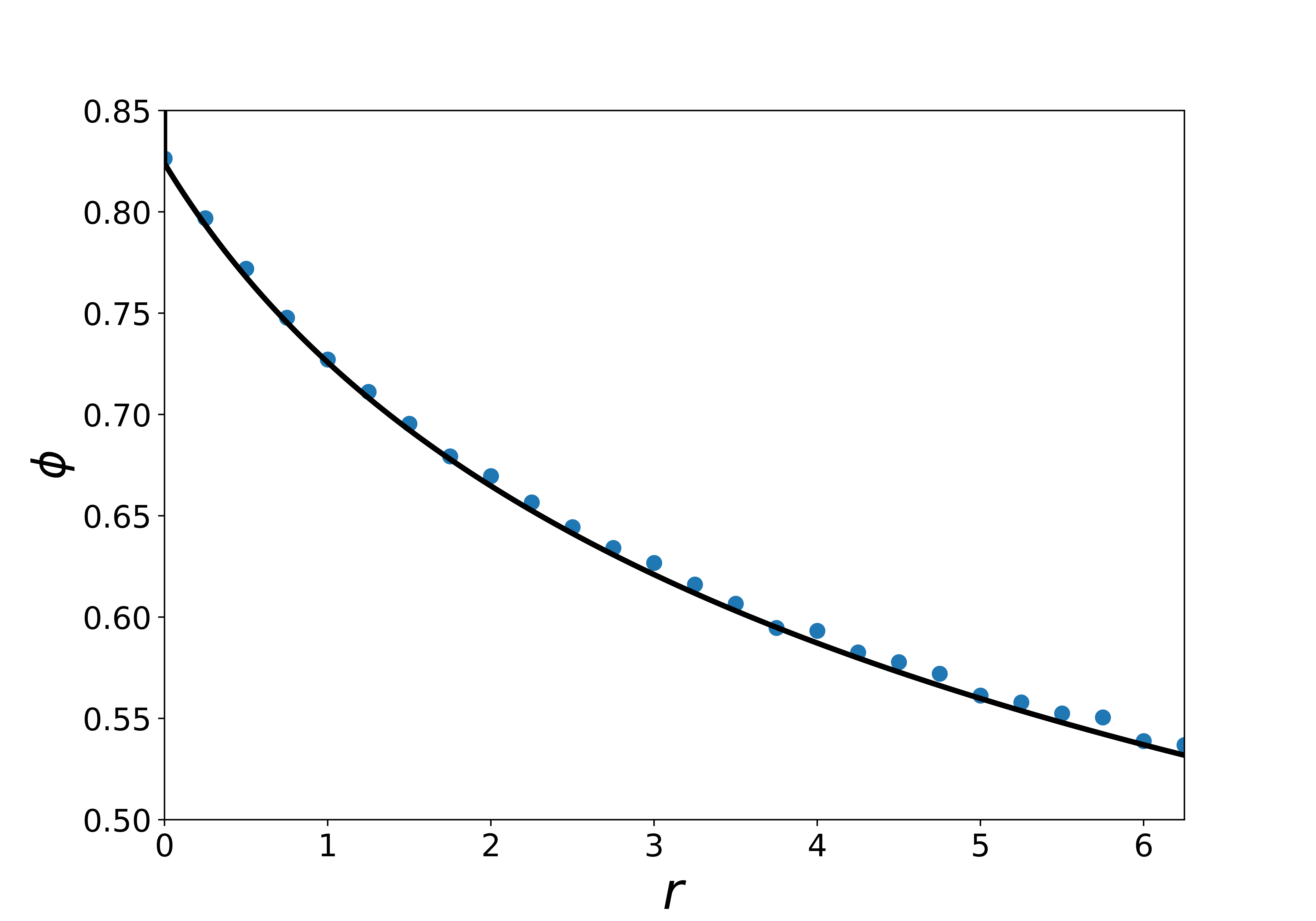}
		\caption{Fraction of surviving species $\phi$ as a function of the in-row correlations $r$. We have fixed $\gamma=\Gamma=0$, $\mu=-1$ and $\sigma = 0.75$. The dots mark the numerical simulations ($N=200$ species, averages over $100$ realisations of the interaction matrix for each data point), produced using a variation of the Runge--Kutta method of fourth order, and the Dormand–Prince method \cite{dormand1980family}. This is implemented in Python using the package ODEint \cite{ahnert2011odeint}. The line shows the predictions from the theory.}
		\label{fig:frac-r}
	\end{figure}
	
	\subsection{Transpose row-column correlations}
\begin{figure}[]
		\centering
		\includegraphics[width=0.9\linewidth]{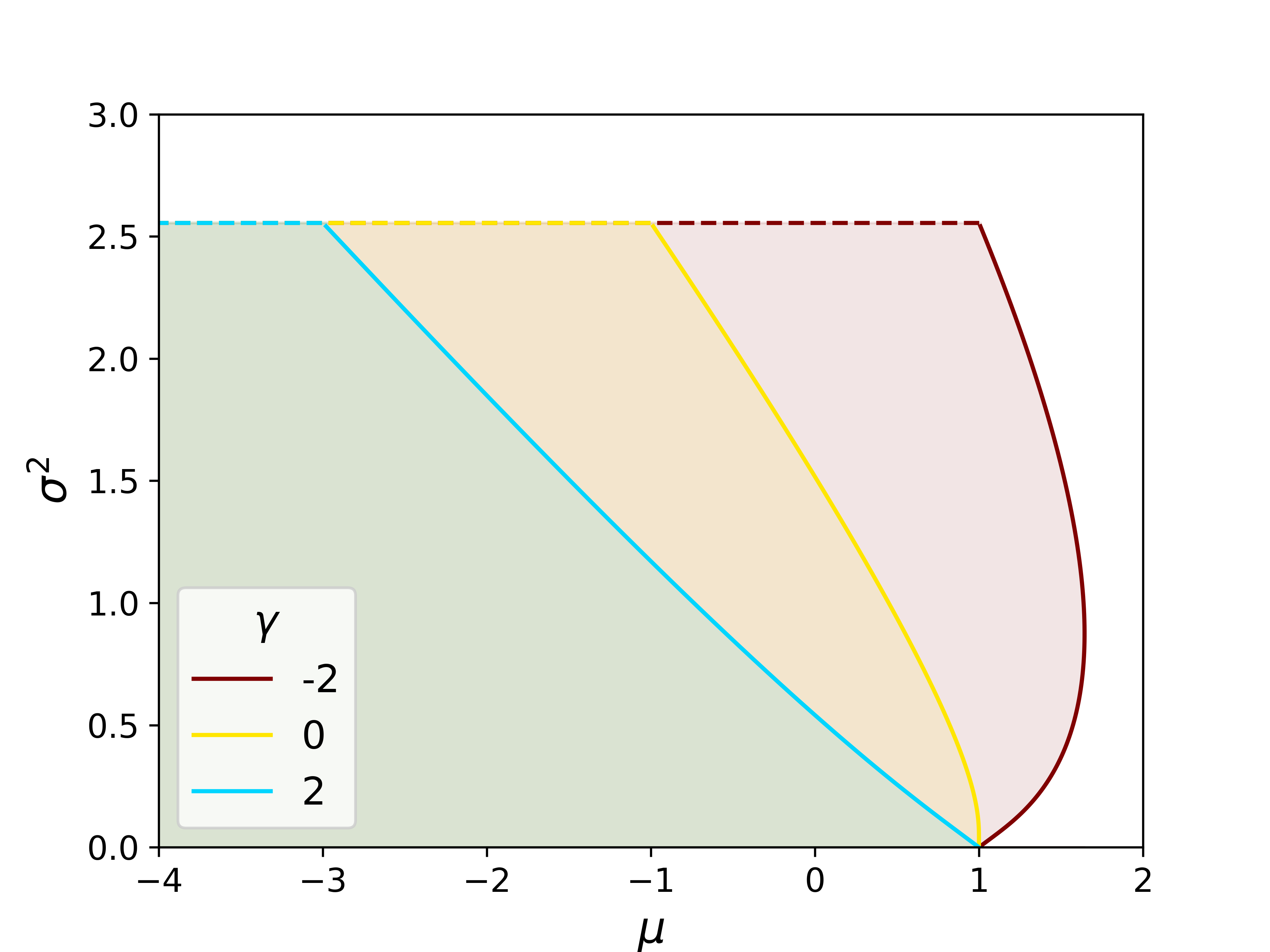}
		\caption{Stability diagram for different values of $\gamma$ at $\Gamma=0 $ and $r=1$. The dashed line represents the transition to linear instability and the solid line indicates the $M\rightarrow \infty$ transition. The area underneath and to the left of the curve is the stable region. It can be seen that higher values of $\gamma$ make the system more likely to diverge but have no impact on the transition to linear instability. In this plot, r is set to 1. One notes that the regions of stability are overlapping.}
		\label{fig:stab-lg}
	\end{figure}

	Fig.~\ref{fig:stab-lg} summarises the effect on stability of varying the degree of correlation between elements $\alpha_{ij}$ and $\alpha_{ki}$, controlled by $\gamma$ in Eq.~(\ref{eq:new-corr}).
	
\begin{figure}
		\centering
		\includegraphics[width=\linewidth]{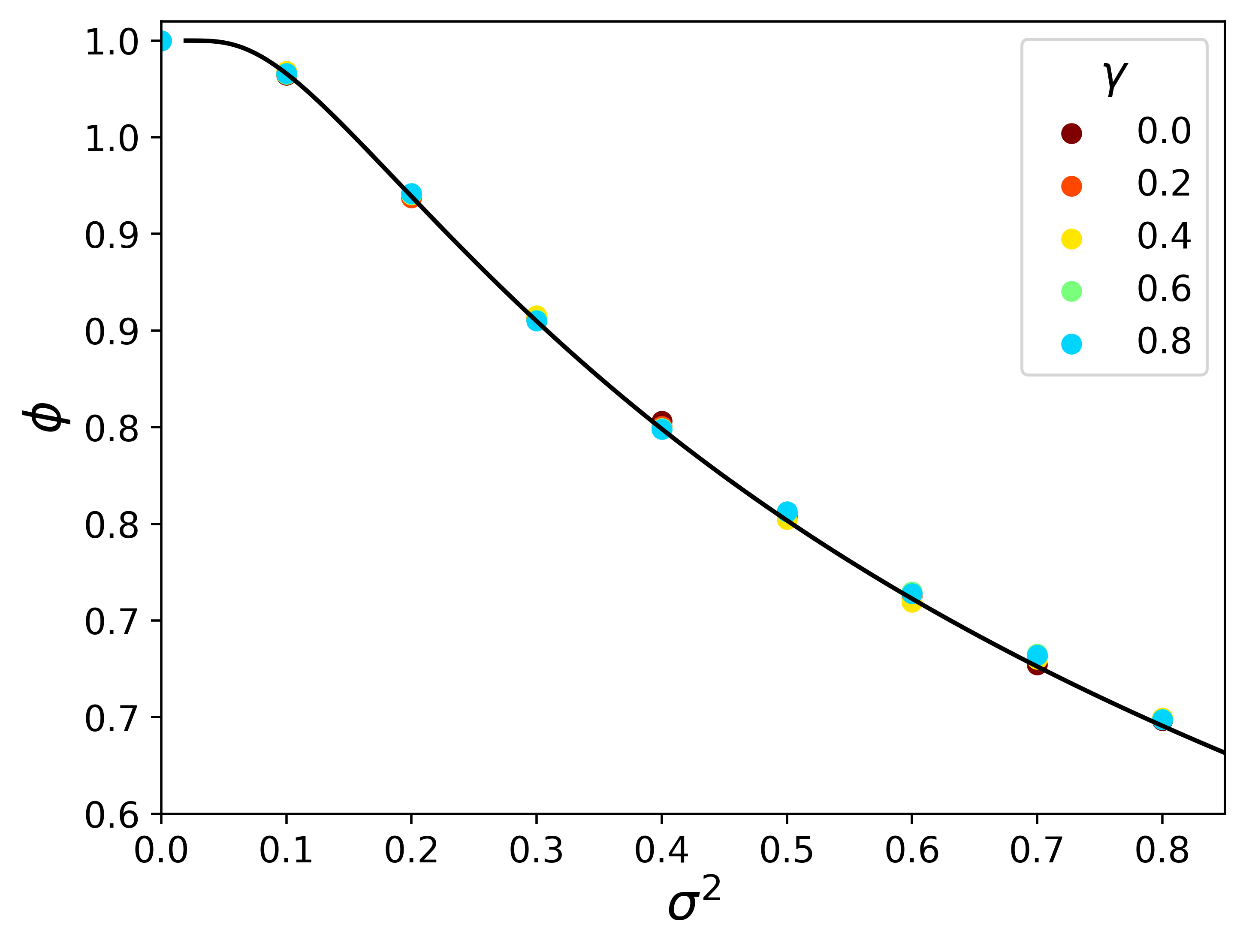}
		\caption{Fraction of surviving species, $\phi$ as a function of the variance of interaction coefficients. Data is shown for transpose row-column correlations (quantified by $\gamma$), at fixed $\Gamma=0$, $r=1$ and $\mu=-1$. The dots mark results from numerical integration of the gLVE ($N=800$ species, averages over $200$ samples of the interaction matrix). The line shows the predictions from the theory (these do not depend on $\gamma$). }
		\label{fig:frac-lg}
	\end{figure}

\begin{figure}
		\centering
		\includegraphics[width=\linewidth]{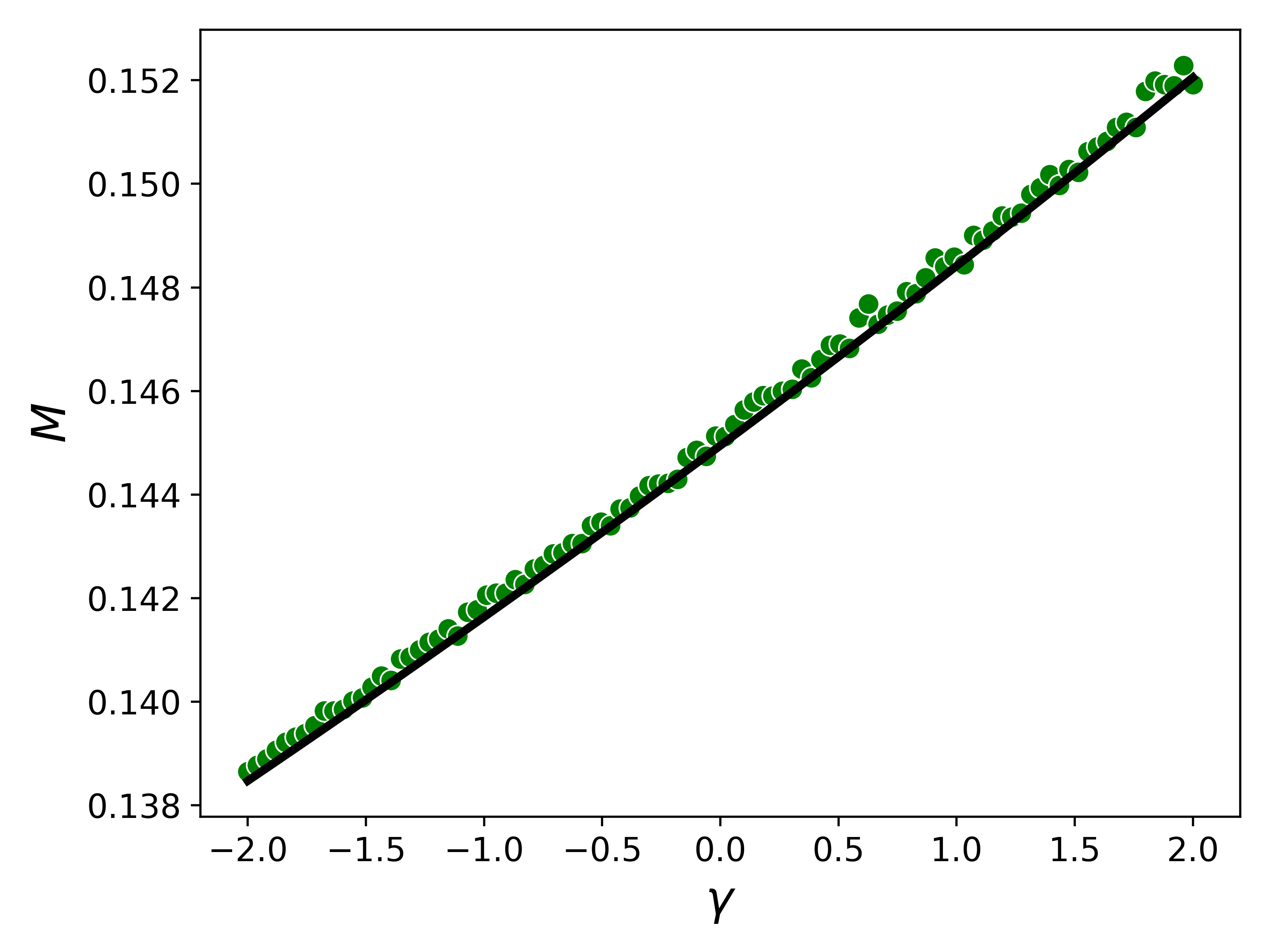}
		\caption{Mean abundance as a function of transpose row-column correlations. Parameters are fixed to $r=1$, $\Gamma = 0.5$, $\sigma = 0.4$ and $\mu = -6$. The dots are results from numerical integration of the gLVE ($N=600$, averaged over 100 samples of the interaction matrix), the line shows the predictions from the theory.}
		\label{fig:lg_M}
	\end{figure}

	For the ${M}\to \infty$ transition, shown by the solid line, we can see that increasing $\gamma$ reduces the stable region. Thus, transpose row-column correlations are destabilising with respect to the divergence of abundances. These results are consistent with \cite{baron2021eigenvalues}, where it was found in a linear system that increasing $\gamma$ also promotes divergence. In contrast with in-row correlations, the value of $\gamma$ has no effect on linear stability (the dotted line in Fig.~\ref{fig:stab-lg} remains the same for all values of $\gamma$). As seen in Eq.~(\ref{eq:opper1}), the onset of linear instability relates to the fraction of surviving species $\phi$. Consistent with the observation that transpose row-column correlations do not alter linear stability, Fig.~\ref{fig:frac-lg} shows that $\phi$ is not affected by varying $\gamma$. However, the mean abundance changes as transpose row-column correlations are varied, as shown in Fig.~\ref{fig:lg_M}. We have also verified these observations by further numerical studies. These are described in Sec.~\ref{sec:num_lg} of the Appendix.

	\subsection{In-Column Correlations}
	\begin{figure}
		\centering
		\includegraphics[width = \linewidth]{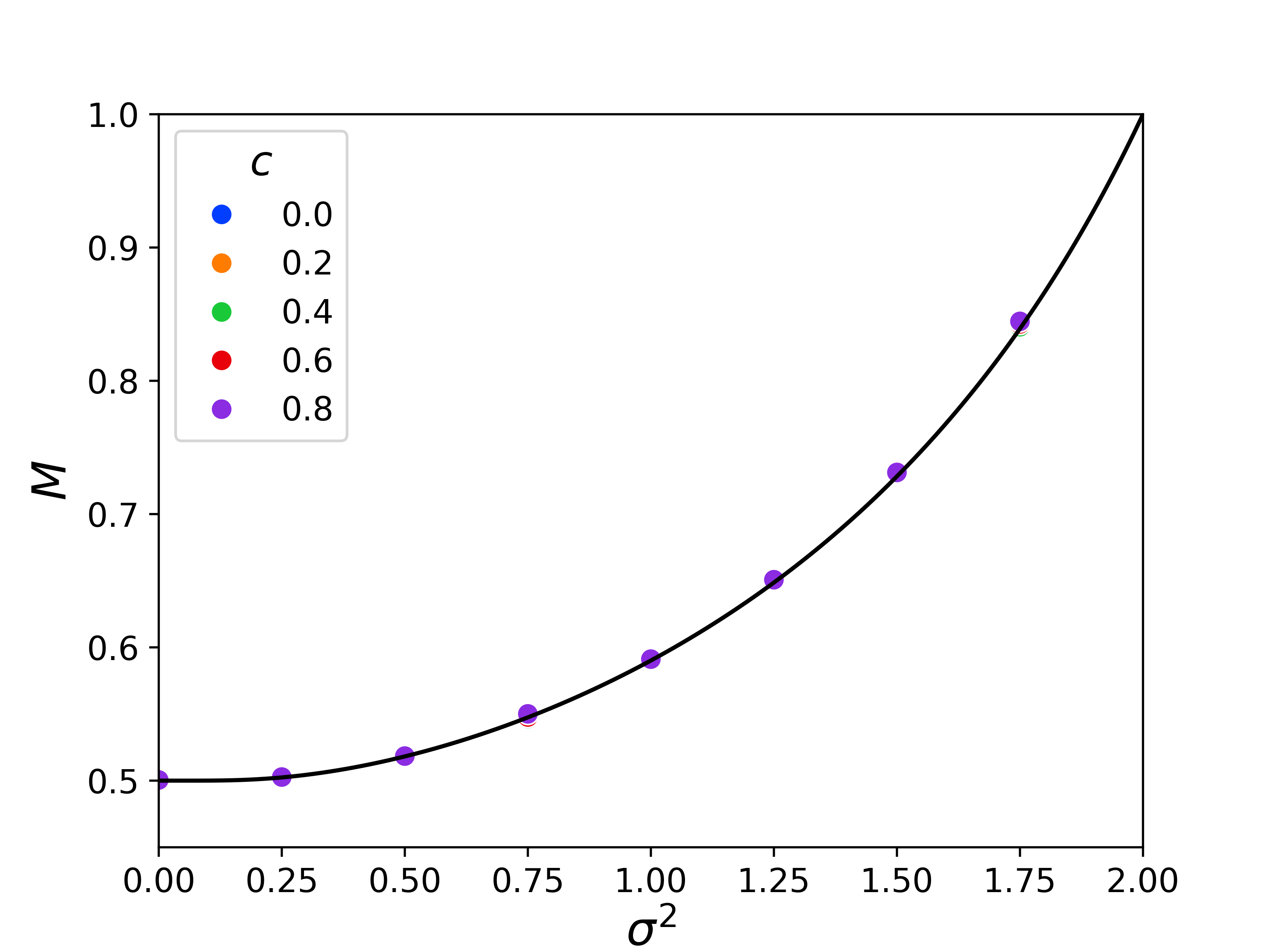}
		\caption{Mean abundance as a function of the variance of interactions for different in-column correlations (determined by $c$). Data is for fixed $r~=~\gamma~=~\Gamma~=~0$ and $\mu=-1$. The dots are results from numerical integration of the gLVE ($N=800$, averaged over $200$ samples of the interaction matrix), the lines are the predictions from the theory.}
		\label{fig:abund-c}
	\end{figure}
	In-column correlations, as defined in Eq.~(\ref{eq:new-corr}) and quantified by the parameter $c$, do not enter into the effective single-species process in Eq.~(\ref{eq:cfs}), nor any of the subsequent analysis. Therefore, we expect the value of $c$ not to have an effect on stability. This is also confirmed in numerical experiments integrating the gLVEs. As shown in Fig. \ref{fig:abund-c} for example, varying $c$ does not affect the mean abundance per species.

	\subsection{Combined effects of the different correlations}
\begin{figure}
		\centering
		\includegraphics[width = 0.88\linewidth]{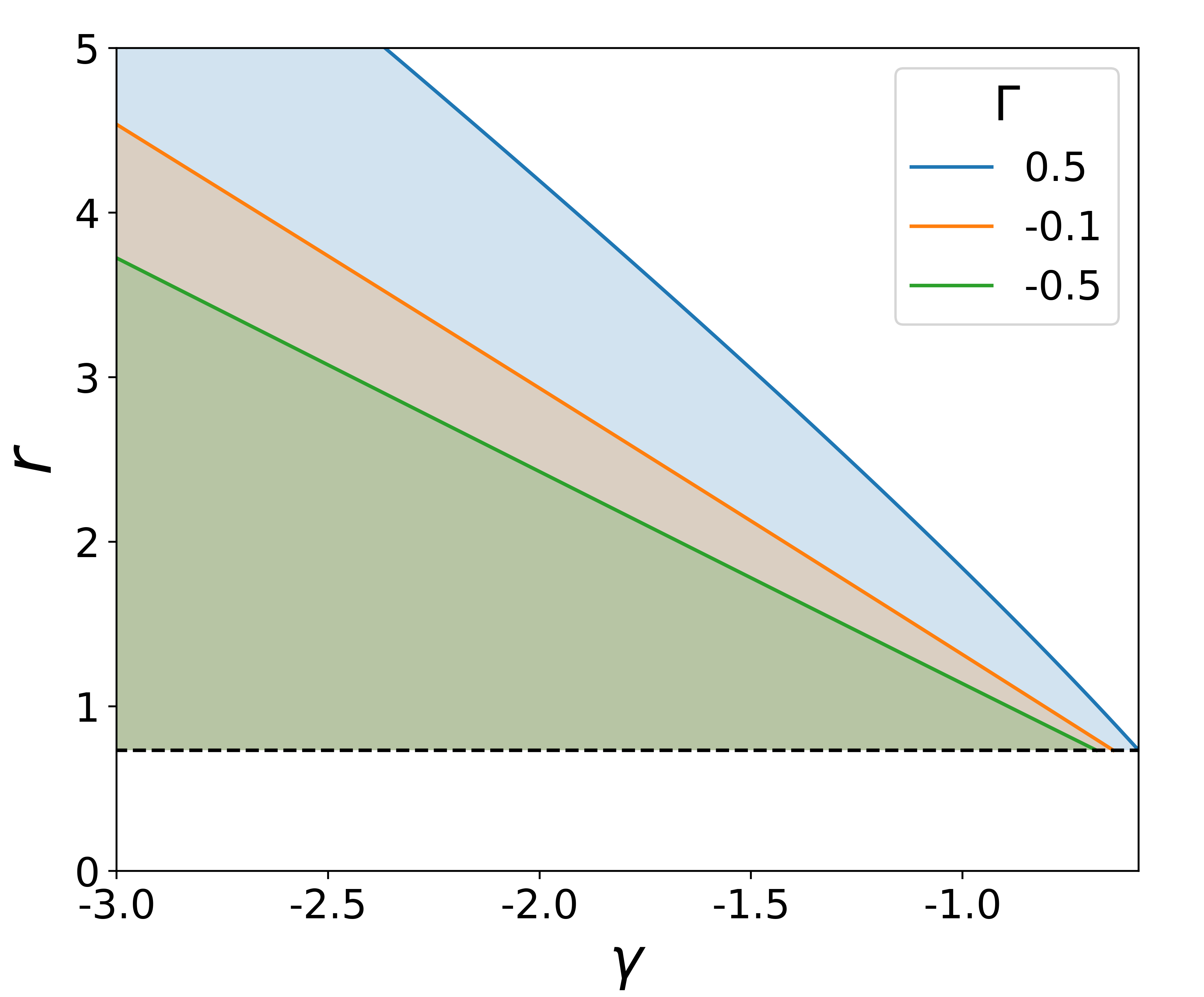}
		\caption{Stability diagram of the gLVE system with simultaneous cross-diagonal, in-row and transpose row-column correlations (parametrised by $\Gamma, r$ and $\gamma$ respectively). We fix $\mu=-0.1$ and set $\sigma$ such that $\sigma^2(1+\Gamma)^2 = 2.2$. The solid lines indicate the $M\to \infty$ transition and the dashed line shows the onset of the linear instability. For each fixed $\Gamma$, the system is stable to the left of the corresponding solid line and above the dashed line.}
		\label{fig:corr_all}
	\end{figure}

	We now give an example illustrating the behaviour of the system if both in-row correlations (parametrised by $r$) and transpose row-column correlations (parametrised by $\gamma$) are present at the same time. 
	Figure~\ref{fig:corr_all} shows the onset of the linear instability (dashed line) and the onset of diverging mean abundance (solid line). These are both shown for different values of the cross-diagonal correlations parameter $\Gamma$.  The stable region for each $\Gamma$ is between the dashed and solid lines. In the figure we have fixed $\mu=-0.1$ and we set $\sigma$ such that $\sigma^2(1+\Gamma)=2.2$. This means that the system is unstable in the absence of generalised correlations [for $r=0$, $\gamma=0$ the system is linearly unstable above $\sigma^2=2/(1+\Gamma)^2$, see Sec.~\ref{sec:lin_instab} and references \cite{bunin2016interaction,bunin2017,galla2018dynamically}].

	Eq.~(\ref{eq:opper1}) [together with Eq.~(\ref{eq:phi})] indicates that the onset of the linear instability does not depend on $\Gamma$ for fixed $\sigma^2(1+\Gamma)$. Additionally, we conclude from Eq.~(\ref{eq:opper2}) that the value of $\gamma$ is irrelevant as well. The dashed line (linear instability) in Fig.~\ref{fig:corr_all} is therefore horizontal and valid for all $\Gamma$.
	
	We conclude from the figure that sufficiently negative transpose row-column correlations (parametrised by $\gamma$), combined with sufficient in row-correlations ($r>0$), can make an otherwise unstable system stable. At the same, time in-row correlations also promote diverging abundances if $r$ is too large. 
	
\begin{figure}
		\centering
		\includegraphics[width=\linewidth]{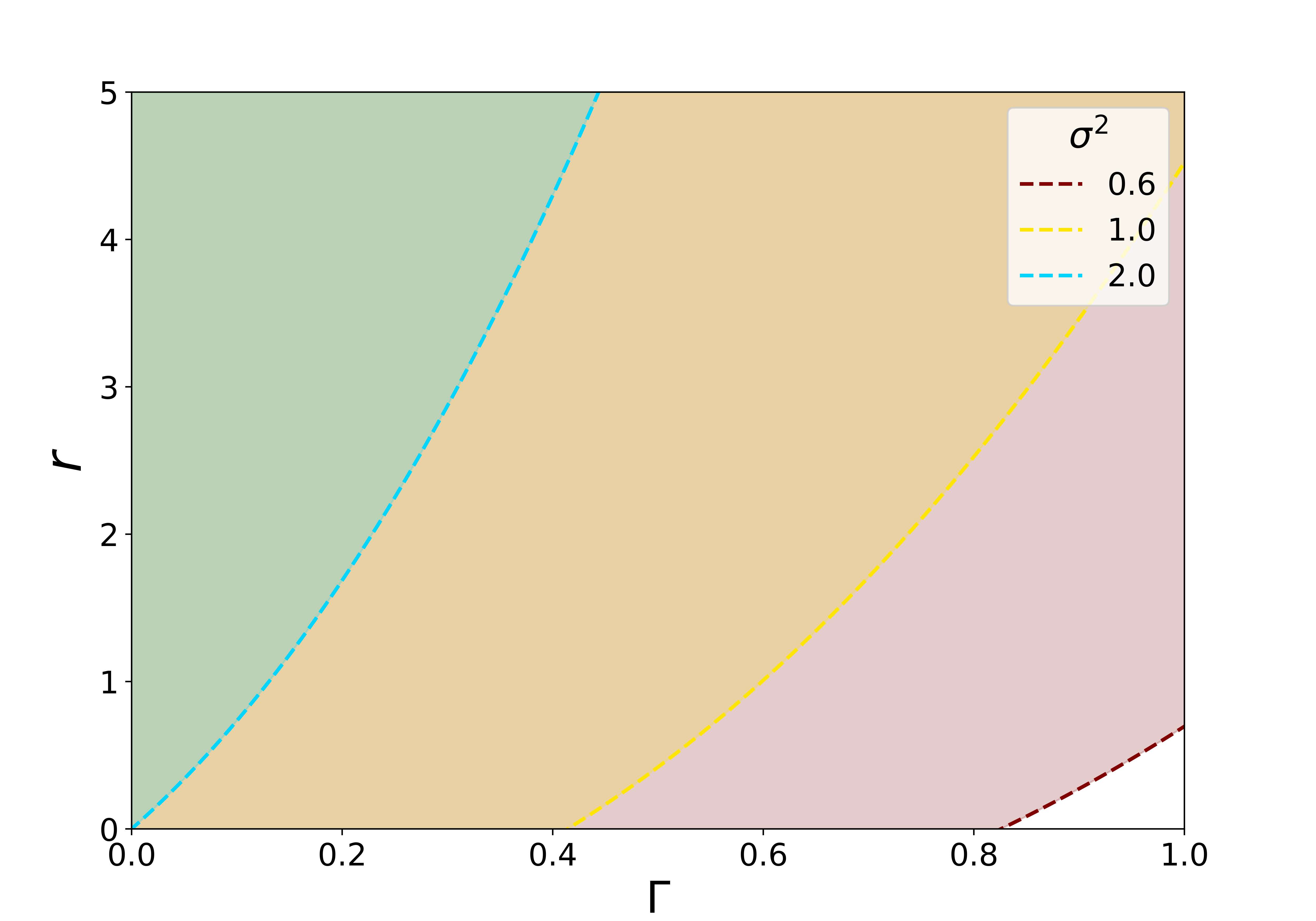}
		\caption{Stability diagram of the linear instability transition in the gLVE system with simultaneous cross-diagonal and in-row correlations (parametrised by $\Gamma$ and $r$ respectively). We fix $\mu=-8$ and vary $\sigma^2$. The solid lines indicate the onset of linear instability. For each fixed $\sigma^2$, the system is stable to the shaded left of the corresponding solid line and the regions are overlapping. This shows how increasing in-row correlations and decreasing cross-correlations make the system more stable. Additionally, the effect of in-row correlations have a diminishing effect on stability at larger values of of $\sigma^2$.}
		\label{fig:opper-Gamma-r}
	\end{figure}

	The onset of the linear instability is only affected by cross-correlations (parametrised by $\Gamma$) and in-row correlations (parametrised by $r$). Fig.~\ref{fig:opper-Gamma-r} illustrates the effects of these correlations. Increasing $\Gamma$ and/or $r$ increases the stability of the system. For larger variation among interaction coefficients (higher values of $\sigma^2$) the stabilising effect of $r$ is diminished.
	
	\section{Discussion of the effects of generalised (single-index) correlations}\label{sec:intuition}
	We have seen that in-row correlations (parameterised by $r$) can increase the amount of effective noise in the system, lead to a greater number of extinctions and promote stability in the surviving community. Our analysis has also shown that transpose row-column correlations (parameterised by $\gamma$) lead to additional memory/feedback effects, and suppress the $M \to \infty$ transition, and that the in-column correlations play no role (at least in the limit $N \to \infty$). 
	
	Below, we provide a more intuitive picture for the reason behind these effects.
	
	\subsection{Effects of generalised correlations on the $M\to\infty$ transition}\label{sec:effM}
	
	\begin{figure}
		\centering
		\hspace{-1em}\includegraphics[width = 1\linewidth]{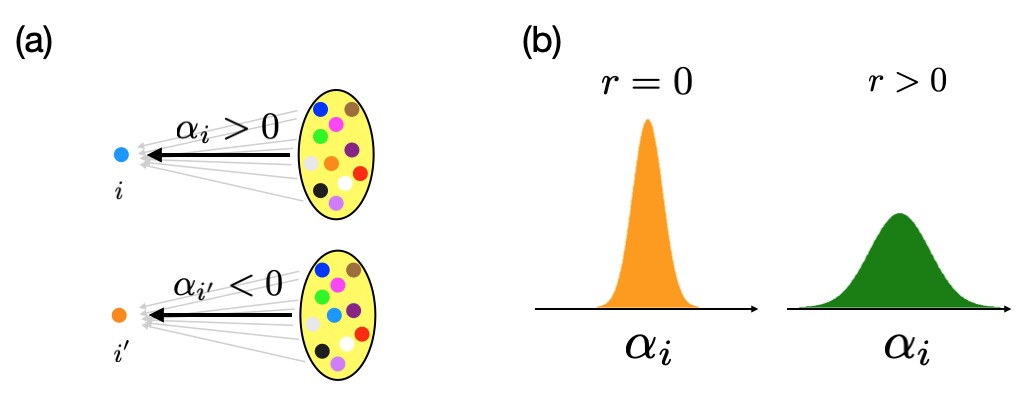}
		\caption{Panel (a): Illustration of the effects $\alpha_i$ and $\alpha_{i'}$ of the pool of species on species $i$ and $i'$. In the example, we have $\alpha_i>0$ and $\alpha_{i'}<0$. Panel (b): Distribution of the random variables $\alpha_i$ for $r=0$ (left) and $r>0$ (right). The $\alpha_i$ are random numbers of mean zero (assuming $\mu=0$) and with variance $(1+r)\sigma^2$ [Eq.~(\ref{eq:var_alpha_i})]. Therefore positive values of $r$ increase the spread of the $\alpha_i$, leading to increasing polarisation of the effects on different species. This polarisation promotes both the divergence of abundances, and an increased fraction of extinct species.}
		\label{fig:interpret_M_r}
	\end{figure}
	\begin{figure}
		\centering
		\includegraphics[width = 1\linewidth]{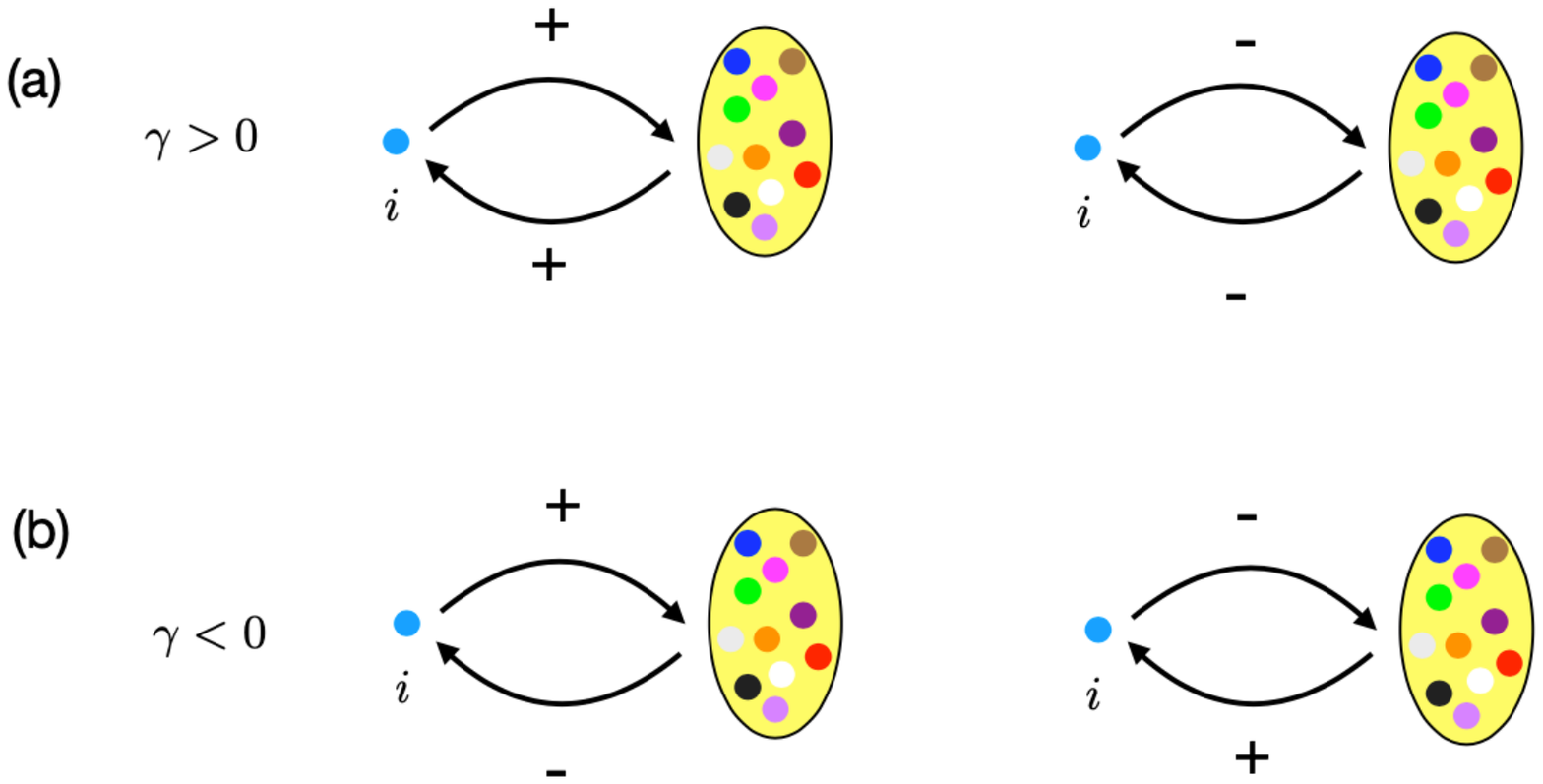}
		\caption{Illustration of the effects of transpose row-column correlations in the transition to divergent abundances.  Transpose row-column correlations modify correlations between the total effect $\beta_i$ a fixed species $i$ has on the bulk of remaining species and the effect $\alpha_i$ that those species have on the focal species $i$. For $\gamma>0$ this promotes positive feedback loops of one species upon itself as illustrated in panel (a).  For example species $i$ may have a positive effect on the bulk, and the bulk also affects the growth of species $i$ positively [panel (a), left]. Alternatively, species $i$ may suppress the growth of species in the remaining pool, but the effect of the pool on $i$ is also negative [panel (a), right]. Either way the feedback loop from species $i$ upon itself is positive, promoting the divergence of abundance. Values $\gamma<0$ promote auto-repressive feedback loops [panel (b)], favouring stability.}
		\label{fig:interpret_M_gamma}
	\end{figure}
	\subsubsection{In-row correlations}
	We have seen that in-row correlations (parametrized by $r>0$), promote the transition to diverging abundances (Fig.~\ref{fig:stab-r}) and reduce the survival fraction (Fig.~\ref{fig:frac-r}). We can provide some intuition by looking at the total strength with which species $i$ is affected (collectively) by all other species. More precisely, we consider $\alpha_i=\sum_{j\neq i}\alpha_{ij}$, assuming that this object, at least approximately, quantifies the overall effect of all species $j\neq i$ on species $i$. 
	
	We have
	\BE
	\mbox{Var}(\alpha_i)&=&\sum_{j\neq i, k\neq i}\overline{z_{ij}z_{ik}} \nonumber \\
	&=&(1+r)\sigma^2 +{\cal O}(N^{-1}).\label{eq:var_alpha_i}
	\EE
	Thus, for $r=0$ the $\alpha_i$ are comparatively tightly distributed around the mean. However, the variance of the $\alpha_i$ increases with $r$. This in turn means that there will be a wider variation of the effects of the system as a whole on individual species, as illustrated for $\mu=0$ in Fig.~\ref{fig:interpret_M_r}. Some species $i$ will experience predominantly positive couplings ($\alpha_i$ positive), and some other species experience mostly negative interactions with the remaining species ($\alpha_i$ negative). The abundances of species with high values of $\alpha_i$ grow, and species with very low $\alpha_i$ are prone to going extinct. Increased in-row correlations promote this polarisation, leading to an increasing fraction of extinct species (Fig.~\ref{fig:frac-r}), and an increased tendency to diverging mean abundance (Fig.~\ref{fig:stab-r}).
	
	\subsubsection{In-column correlations}
	Similar to the quantity $\alpha_i$ above, we define $\beta_i=\sum_{j\neq i} \alpha_{ji}$. Thus, $\beta_i$ quantifies the total effect species $i$ has on all other species. We then have
	\BE
	\mbox{Var}(\beta_i)&=&\sum_{j\neq i, k\neq i}\overline{z_{ji}z_{ki}} \nonumber \\
	&=&(1+c)\sigma^2 +{\cal O}(N^{-1}).
	\EE
	Values of $c>0$ therefore give rise to a greater `polarisation' of the effect of individual species on the pool of the remaining species. Some species $i$ will have a net negative on the pool, and the effect of another species $i'$ on the pool might be positive. For increased $c$, any species in the pool is therefore subject to a wider range of positive and negative influences. Nonetheless, these positive and negative influences on any fixed species average out, and the mean effect on any species in the pool does not change with $c$. In-row correlations therefore do not affect the onset of diverging mean abundance.

	\subsubsection{Transpose row-column correlations}
	We have
	\BE\mbox{Cov}(\alpha_i,\beta_i)&=&\sum_{j\neq i, k\neq i} \overline{z_{ij}z_{ki}} \nonumber \\
	&=& (\Gamma +\gamma)\sigma^2 +{\cal O}(N^{-1}).
	\EE
	Therefore $\alpha_i$ and $\beta_i$ will be more positively correlated when $\gamma>0$. If the effect species $i$ has on the remaining species is positive, then the effect of those other species on $i$ is also likely to be positive. Similarly, if $i$ acts negatively on the pool of species, then the effect of the pool on $i$ will also tend to be be negative. Either way, this promotes positive feedback of species $i$ on itself, as illustrated in Fig.~\ref{fig:interpret_M_gamma}(a). This  promotes the growth of species $i$ (and similarly for all other species). Therefore, increased positive transpose row-column correlations will promote positive feedback and, consequently, the divergence of the mean abundance (Fig.~\ref{fig:stab-lg}).
	
	Negative values of $\gamma$ on the other hand result in more negative correlations between $\alpha_i$ and $\beta_i$.  Negative values of $\gamma$ promote negative feedback loops of species $i$ on itself [Fig.~\ref{fig:interpret_M_gamma}(b)], and therefore limits the overall growth. This leads to an increased region of stability, as seen in Fig.~\ref{fig:stab-lg}.
	
	\subsection{Effects of generalised correlations on linear instability}\label{sec:eff_lin_instab}
	We have seen that the linear instability is not affected by in-column correlations or transpose row-column correlations (parametrised by $c$ and $\gamma$ respectively). However, increased positive in-row correlations (given by $r$) make the system more stable against small perturbations (Fig.~\ref{fig:stab-r}).
	
	The linear instability is a result of (i) the degree of variation of interaction coefficients and (ii) the extinction dynamics in the gLVE system, leading to a community of survivors consisting of a fraction $\phi$ of the original pool of $N$ species.

	A simple argument for the effects of in-row correlations on the onset of linear instability can therefore be made by referring to the May bound on stability. An equilibrium of a complex system is stable only if its `complexity' remains below a certain bound \cite{may71}. In our case, the complexity of the system is given by the product of its size (number of species) and the variation of interaction coefficients. For the surviving community this is proportional to $\phi \sigma^2$. In-line with this reasoning we find that the onset of instability occurs when $\sigma^2\phi=(1+\Gamma)^{-2}$ [Eq.~(\ref{eq:opper1})]. At the same time, we have found that in-row correlations reduce the size of the surviving community, keeping all other parameters fixed (see Fig.~\ref{fig:frac-r}). This in turn means that the system can maintain stability for a higher variation $\sigma^2$ of the interaction coefficients than without in-row correlations. Thus, the linear instability line in Fig.~\ref{fig:stab-r} is pushed upwards for increasing in-row correlations. Mathematically, this can also be seen from Eq.~(\ref{eq:opper2}). For $r=0$ the  linear instability occurs at $\Delta_c=0$, implying that exactly half of the initial pool of species survives ($\phi=w_0(\Delta_c)=1/2$). When $r>0$, we find $\Delta_c<0$, and thus the fraction of survivors is smaller than $1/2$ at the point where the system becomes linearly unstable. We discuss this matter in greater detail in Appendix.~\ref{app:lin_instab}.
	
	We also note that the fraction of surviving species is not affected by either transpose row-column correlations or in-column correlations, which is why these correlations have no effect on the onset of linear instability.

	\section{Conclusions}
	\label{sec:disc}
	
	In summary, we have used dynamical mean field theory to study the most general Lotka--Volterra system with random interactions that do no give statistical preference to any species over another. This system is defined by an interaction matrix with correlations not only between diagonally opposed entries, but also between entries sharing only a single index. That is, we study the effects of correlations between entries within rows, within columns, and between entries in transpose pairs of rows and columns. 
	
	We demonstrated using DMFT that positive in-row correlations give rise to a greater effective degree of noise in the system. Also, correlations between transpose pairs of rows and columns gave rise to additional system-wide feedback effects. These observations are entirely general, and ought to apply to any dynamical model involving quenched disordered interactions with generalised (single-index) correlations. On a technical note, we also observed that in the presence of single-index correlations, the DMFT analysis using a path-integral approach was more transparent than the cavity method, whereas normally the two approaches would be equally useful (in densely interacting systems).
	
	Negative correlations between diagonally opposed elements are known to promote stability \cite{bunin2016interaction, bunin2017, galla2018dynamically} in the Lotka-Volterra model, and other systems \cite{galla2006, kuczala2016eigenvalue}. Our analysis has similarly allowed us to identify the effect of the additional single-index correlations on stability in the gLVEs. We find that in-column correlations do not affect the stability of the system nor the composition of the community of surviving species. In-row correlations promote the divergence of abundances, but reduce the size of the surviving community and therefore make the system more stable against small perturbations. Transpose row-column correlations also promote the divergence of abundances, but have no effect on the size of the surviving community nor linear stability. 
	
	On the subject of direction for further investigation, it is known that the statistics of the interactions between extant species can be very different from the ones in the initial pool of all species \cite{bunin2016interaction, baron_et_al_prl}. In particular, single-index correlations are seen in the interaction matrix of the surviving community, even if they are not present in the original community. We therefore expect that the extinction dynamics modifies these generalised correlations if they are present from the beginning. It would be interesting to investigate what exactly these modifications are. Further, the relationship between surviving interaction statistics and network structure has also been investigated \cite{poley2, park, aguirre}. In light of our results demonstrating the role of in-row correlations on the extinction process, it would therefore be interesting to explore further how sensitive the surviving network structure is to such intricacies. That is, it is still left to understand what kinds of subtle `fingerprints' of the history of an ecosystem it may be possible to detect in the long term behaviour of the community.

	\section*{Acknowledgements}
	TG acknowledges partial financial support from the Agencia Estatal de Investigaci\'on and Fondo Europeo de Desarrollo Regional (FEDER, UE) under project APASOS (PID2021-122256NB-C21, PID2021-122256NB-C22), and the Maria de Maeztu programme for Units of Excellence, CEX2021-001164-M funded by  MCIN/AEI/10.13039/501100011033. JWB was supported by grants from the Simons Foundation (\# 454935 Giulio Biroli). 
	\newpage
	\begin{appendix}\onecolumngrid
		\section{Full Form of Generating Functional}
		\label{sec:full_gen}
		The generating-functional calculation follows the lines of \cite{galla2018dynamically,galla_notes} closely, subject to relatively minor and straightforward modifications required to account for the generalised correlations. Because of this, we do not repeat the calculation in all detail here. Instead we provide, in brief form, some intermediate steps to assist readers who would like to re-construct the full calculation
		
		The first main steps of the procedure are (i) the construction of the generating functional for the gLVE system, and (ii) carrying out the disorder average of the generating functional. The disorder-averaged generating functional can then be written in saddle-point form as follows
		\begin{equation} 
			\overline{Z[\boldpsi]}
			= \int D[{M},{{C}},{{L}},{{K}},{P},{\hat{M}},{{\hat{C}}},{{\hat{L}}},{{\hat{K}}},{\hat{P}}] \,\exp{N(\Psi+\Phi+\Omega+ \mathcal{O}(N^{-1}))},
			\label{eqn:disorder_avg_OP}
		\end{equation}
		where $\Psi$ contains the macroscopic order parameters and their conjugate variables, $\Phi$ are the terms resulting from the disorder average and $\Omega$ describes the resulting mean field dynamics. The notation is broadly as in \cite{galla_notes}.
		
		We here provide the different terms in the exponential. We have
		\BE
		\Psi &= &i \int dt \left[{\hat{M}}(t){M}(t) +  {\hat{P}}(t) {P}(t)\right] \nonumber \\
		&& + i\int dt \int dt'\left[{\hat{C}}(t,t')) {C}(t,t')  +  {\hat{K}}(t,t')){K}(t,t') + {\hat{L}}(t,t')){L}(t,t')\right].
		\EE

		Further, $\Phi$ takes the form
		\BE\label{eq:phi}
		\Phi &= &-\frac{1}{2}\sigma^2\int dt\int dt'\biggl[L(t,t')C(t,t') + \Gamma K(t,t')K(t',t) + rL(t,t')M(t)M(t')\biggl.\nonumber \\
		&&\biggl. - cC(t,t')P(t)P(t') - 2\gamma i K(t',t)M(t)P(t')\biggr]-\mu\int dt M(t)P(t).
		\EE
		Finally, we have
		\BE    
		\Omega &=&  \log\Bigg\{\int D[x,\hat x] p(x(0) \exp(i\int dt \psi(t)x(t))\exp\left(i \int dt \hat x(t)\left(\frac{\dot x(t)}{x(t)}-(1-x(t))-h(t)\right)\right)\nonumber \\
		&&        \exp\left(-i\int dt \int dt'  {\hat C}(t,t')x(t)x(t') +  {\hat L}(t,t') \hat{x}(t)\hat{x}(t') + {\hat K}(t,t') x(t)\hat{x}(t')\right)\nonumber \\
		&& \exp\left(-i\int dt  {\hat M}(t) x(t) + i {\hat P}(t)\hat{x}(t)\right)\Bigg\}.
		\EE
		The $\hat x_i(t)$ are conjugate variables to the $x_i(t)$, and we have introduced the macroscopic order parameters (for details see again \cite{galla_notes}),
		\begin{equation}
			\begin{split}
				{M}(t) &= \frac{1}{N}\sum_{i}x_{i}(t),\\
				{P}(t) &= i\frac{1}{N}\sum_i\hat{x}_i(t),\\
				{{C}}(t,t') &= \frac{1}{N}\sum_ix_i(t)x_i(t'),\\
				{{K}}(t,t') &= \frac{1}{N}\sum_ix_i(t)\hat{x}_i(t'),\\
				{{L}}(t,t') &= \frac{1}{N}\sum_i\hat{x}_i(t)\hat{x}_i(t').
				\label{eqn: OP_1}
			\end{split}
		\end{equation}
		The variables $\hat C(t,t'), \hat K(t,t'), \hat L(t,t), \hat M(t)$ and $\hat P(t)$ are conjugate fields to the macroscopic order parameters.
		
		We now follow the usual steps \cite{coolen_dyn, Coolen_MG,galla2018dynamically,galla_notes} and extremise the exponent in Eq.~(\ref{eqn:disorder_avg_OP}) with respect to $C(t,t'), K(t,t'), L(t,t'), M(t)$ and $P(t)$. This allows us to express $\hat C(t,t'), \hat K(t,t'), \hat L(t,t), \hat M(t)$ and $\hat P(t)$ in terms of the (unhatted) macroscopic order parameters in the expression for $\Omega$. After further standard manipulations we then obtain the following generating functional for the effective mean-field process
		\BE\label{eq:gf_effective}
		Z_{\text{eff}} &=& \int D[{x},{\hat{x}}]p(x(0)) \exp\left(i\int dt \;\hat{x}(t)\left[\frac{\dot{x}(t)}{x(t)}-1+x(t)-\mu {M}(t)-h(t)\right]\right) \nonumber \\
		&&\exp\Big(-\sigma^2 \int dt\, dt' \Big[\frac{1}{2} \hat{x}(t)\hat{x}(t')({{C}}(t,t')+ r{M}(t){M}(t')) + i{{G}}(t',t)\left[\Gamma x(t) \hat{x}(t')+\gamma \hat{x}(t'){M}(t)\right]\Big]\Big).
		\EE
		Here, $\Gamma$, $r$ and $\gamma$ are the correlations defined in Eq.~(\ref{eq:new-corr}), $\sigma$ and $\mu$ are as in Eq.~(\ref{eq:alpha}). We have written $G(t,t)=-iK(t,t')$. From Eq.~(\ref{eq:gf_effective}) one then obtains the dynamic mean field theory in Eq.~(\ref{eq:eff_proc}) and the self-consistency relations in Eq.~(\ref{eq:cfs}).
		
		\section{Solution Procedure}
		\label{sec:sol_proc}
		Through algebraic manipulation of the relations in Eqs.~(\ref{eqn:op_static}) and (\ref{eqn:delta_def}), we find
		\begin{subequations}\label{eq:sol_proc}
			\begin{gather}
				\chi = w_0 + \frac{w_0^2}{w_2}\frac{\Gamma}{1+r\frac{w_1^2}{w_2}},\\
				\sigma^2 = \frac{w_2}{\left[w_2+\frac{\Gamma w_0}{1+\frac{rw_1^2}{w_2}}\right]^2\left(1+\frac{rw_1^2}{w_2}\right)},\label{eq:sol_sigsq}\\
				\frac{1}{M^*} = \frac{\Delta w_2 \left(1+\frac{rw_1^2}{w_2}\right) - \gamma w_0 w_1}{w_1 \left[w_0\Gamma + w_2\left(1+\frac{rw_1^2}{w_2}\right)\right]} - \mu,\label{eq:sol_1_M}\\
				q = \frac{(M^*)^2w_1^2}{w_2},\\
				\phi = w_0.
			\end{gather}
		\end{subequations}
		To keep these relations compact, we have omitted the argument $\Delta$ of the functions $w_\ell$ on the right ($\ell=0,1,2,3$). These functions are defined in Eq.~(\ref{eqn:delta_def}).

		Eqs.~(\ref{eq:sol_proc}) allow us to calculate the static order parameters as a function of $\sigma^2$ in parametric form, following steps similar to \cite{galla2018dynamically,galla_notes}. To do this, we fix the values of $\mu$, $\Gamma, \gamma$ and $r$, and then obtain $\sigma^2,\, q,\, \chi$ and $M^*$ as functions of $\Delta$. 
		\section{Addtional simulation results}
		\subsection{Numerical Verification of effects of in-row correlations}
		\label{Sec:num_r}
		In this section we provide further evidence for the effects of in-row correlations on the onset of the linear instability and on the divergence of abundances. 
		
		\subsubsection{Linear instability}
		For fixed instances of the interaction matrix we ran the gLVE dynamics for two random initial conditions, and obtained trajectories $x_i(t)$ and $x_i'(t)$. We then measured the relative time-averaged distance between these trajectories as in \cite{sidhomgalla},
		\begin{equation}
			d = \frac{\langle\langle \left(x_i(t)-x_i'(t)\right)^2\rangle_N\rangle_T}{\langle\langle \left(x_i(t)\right)^2\rangle_N\rangle_T}.
			\label{eq:distance}
		\end{equation}
		We have written $\langle x_i^2\rangle_N = N^{-1} \sum_ix_i^2$ (a similar definition applies in the numerator), and $\langle \dots \rangle _T$ denotes a time average in the stationary state.
		
\begin{figure}[htbp]
			\centering
			\begin{minipage}[t]{0.48\linewidth}
				\centering
				\includegraphics[width=\linewidth]{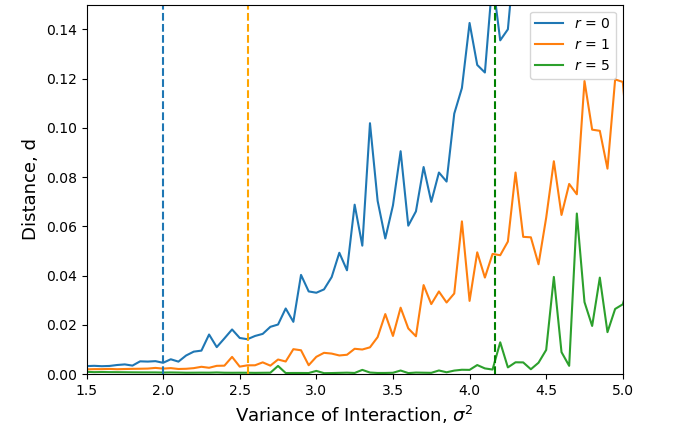}
				\caption{Normalised asymptotic average distance between different trajectories for the same interaction matrix. We show data for different in-row correlations $r$, and fixed $\Gamma=\gamma=0$ and $\mu=-6$. The dashed lines indicate the onset of the linear instability as predicted by the theory. The solid lines are from numerical integration of the gLVE system with $N=800$, averaged over $200$ samples of the interaction matrix.}
				\label{fig:sim-opper-r}
			\end{minipage}
			\hfill
			\begin{minipage}[t]{0.48\linewidth}
				\centering
				\includegraphics[width=\linewidth]{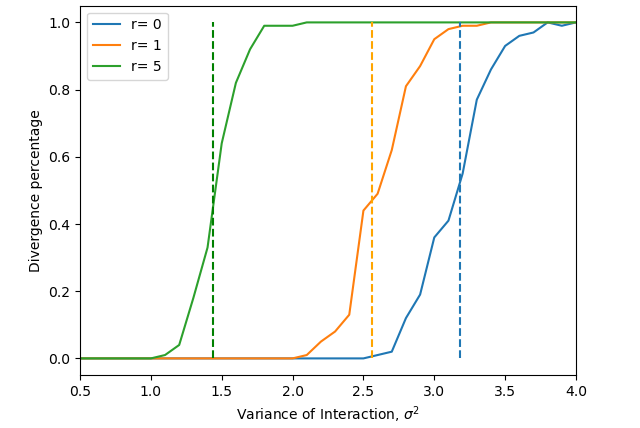}
				\caption{Fraction of samples identified as divergent as a function of $\sigma^2$ for different in-row correlations (parametrised by $r$). The dashed vertical lines mark the predicted onset of diverging abundance from the theory. Data is for $\Gamma=\gamma=0$, $\mu = -1$ and $N=250$. The dynamics was averaged over $200$ samples}
				\label{fig:mifty_sim}
			\end{minipage}
		\end{figure}

		In the stable phase, the system has a unique attracting fixed point for any realisation of the interaction matrix. We thus expect $d=0$ (in the limit $N\to\infty$). Positive values of $d$ indicate that the assumptions of a unique stable fixed point no longer hold. Thus we can use numerical measurements of $d$ to investigate the boundaries of the stable phase. In simulations for finite $N$, the transition from $d=0$ to $d>0$ will never be perfectly sharp, and therefore we do not claim that Fig.~\ref{fig:sim-opper-r} provides quantitative confirmation of the point at which the linear instability sets in. Nonetheless, as seen in the figure the range of $\sigma^2$ for which the distance $d$ remains close to zero grows with $r$. This broadly confirms that increasing in-row correlations delay the onset of the linear instability [see also Fig.~\ref{fig:stab-r}].

		\subsubsection{Diverging mean abundance}
		
		Numerical evidence for the effects of in-row correlations on the divergence of abundances can be found in Fig.~\ref{fig:mifty_sim}. The dashed vertical lines mark the theoretically predicted point at which $M^*\to \infty$. The solid lines were obtained from numerically integrating the gLVE varying $\sigma^2$ for fixed $\Gamma,\,\gamma$ and $r$ until any of the species experiences an increase of more than $10^4$ in a single timestep of the numerical integration, which we then note as a divergence. We set the timestep to 0.01 in our simulations. Data is averaged over $200$ samples of the interaction matrix.
		
		As in the previous figure, the data is subject to finite-size effects. Nonetheless, the simulations broadly confirm the analytical predictions. In particular, increasing the in-row correlation parameter $r$ moves the onset of divergence to smaller values of $\sigma^2$, as also shown in Fig.~\ref{fig:stab-r}.
		
		\subsection{Verification of the effects of transpose row-column correlations}
		\label{sec:num_lg}
		\subsubsection{Linear instability}
\begin{figure}[htbp]
			\centering
			\begin{minipage}[t]{0.48\linewidth}
				\centering
				\includegraphics[width=\linewidth]{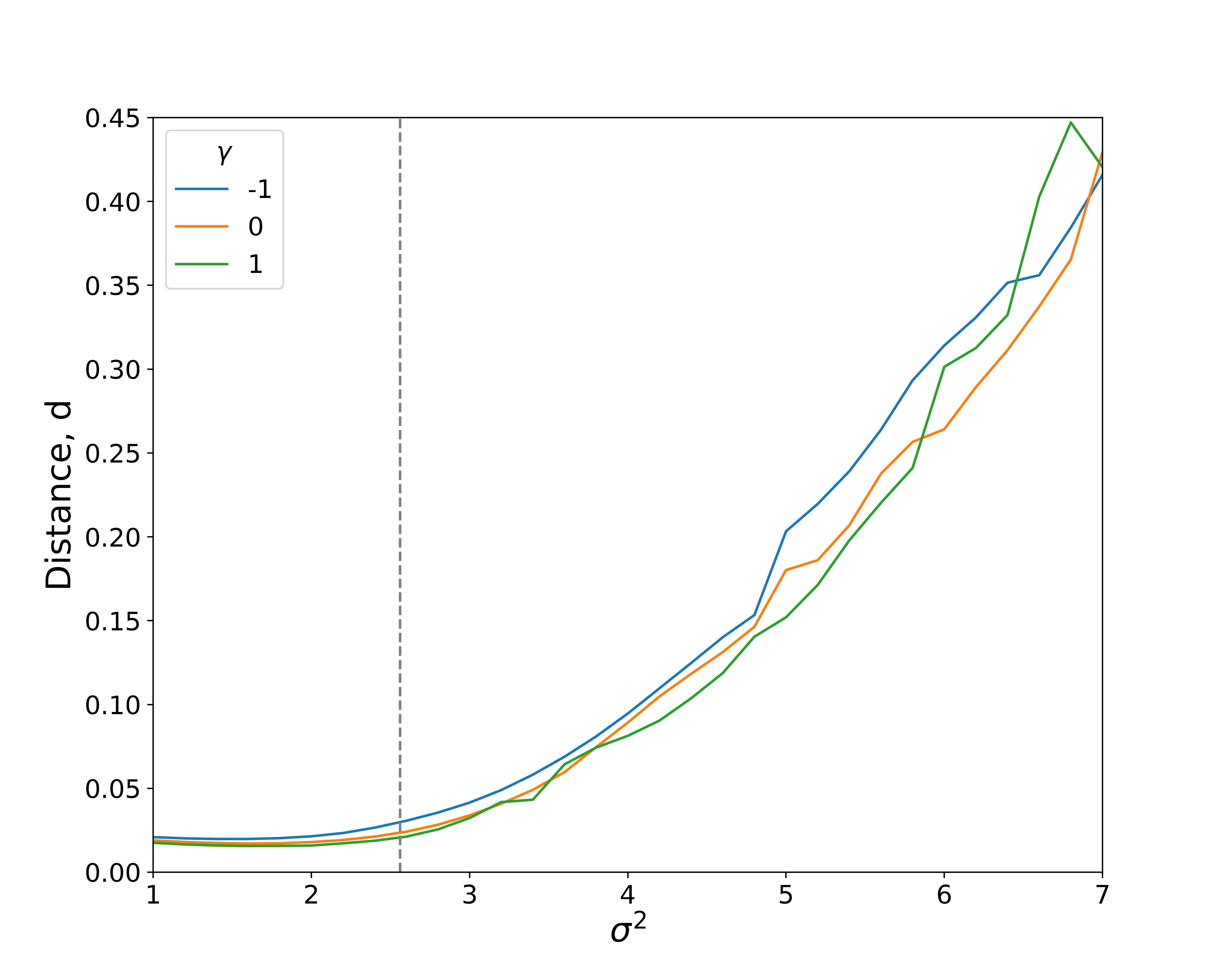}
				\caption{Numerical verification of the independence of the distance $d$ of the linear instability on $\gamma$. Model parameters are $\Gamma = 0$, $r = 1$ and $\mu=-10$. The dashed line indicates the theoretical onset of the linear instability. Data is for $N=300$, averages over 200 realisations are taken.}
				\label{fig:sim-opper-lg}
			\end{minipage}
			\hfill
			\begin{minipage}[t]{0.48\linewidth}
				\centering
				\includegraphics[width=\linewidth]{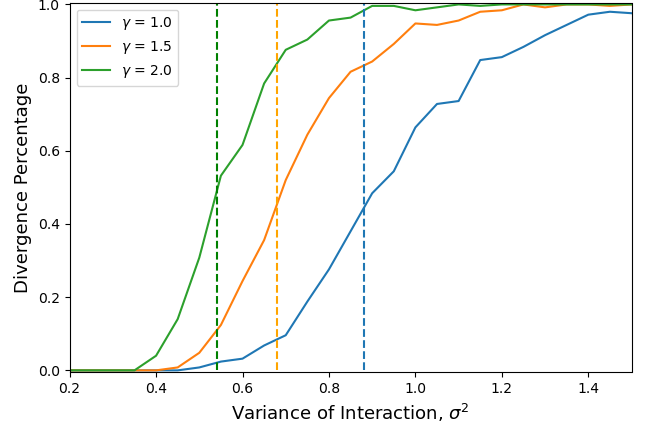}
				\caption{Numerical verification of diverging ecosystems for varied $\gamma$, $\Gamma = 0$, $r = 1$ and $\mu=0$. The dashed vertical lines mark the theoretically predicted point of instability derived from Figure~\ref{fig:stab-lg}. Data is for $N=300$, averages over 200 realisations are taken.}
				\label{fig:mifty_sim-lg}
			\end{minipage}
		\end{figure}

		The theory predicts that the onset of the linear instability is not affected by transpose row-column correlations (parametrised by $\gamma$). The numerical results in Fig.~\ref{fig:sim-opper-lg} are consistent with this. We show the distance $d$, defined in Eq.~(\ref{eq:distance}), as a function of $\sigma^2$ for different values of $\gamma$, at fixed $\Gamma, r$ and $\mu$. The data shows that $d$ is not affected by the transpose row-column correlations and the distance begins to rise noticeably after a critical value of $\sigma^2$, derived from the theory. As in earlier figures, the quantity $d$ is subject to finite-size effects and therefore not strictly zero in the stable phase. The data in Fig.~\ref{fig:sim-opper-lg} is therefore not intended to be a quantitative verification of the onset of the linear instability, but rather as further demonstration of the irrelevance of transpose row-column correlations for the onset of linear instability.
		
		\subsubsection{Onset of diverging abundance}
		We study the effects of transpose row-column correlations on the onset of diverging abundances in Fig.~ \ref{fig:mifty_sim-lg}. The dashed vertical lines mark the point of divergence as predicted by the theory. The solid lines are from numerical integration of the gLVE and show the fraction of samples that diverged. Due to finite size effects it is hard to obtain a quantitative confirmation of the analytical predictions, however, we notice that for all three values of $\gamma$ shown the fraction of divergent samples crosses $50\%$ approximately at the theoretically predicted value. The figure shows that increasing transpose row-column correlations (larger $\gamma$) moves the onset of diverging abundances to smaller values of the variance $\sigma^2$ of interactions, in-line with Fig.~\ref{fig:stab-lg}.

		\section{Further comments on the onset of linear instability}\label{app:lin_instab}
		In Sec.~\ref{sec:eff_lin_instab} of the main paper we have provided a heuristic explanation of the effects of in-row correlations on the onset of linear instability. This argument refers to the size of the community of surviving species, and on the dependence of $\phi$ (the fraction of survivors) on the in-row correlations.
		
		In this section we comment further on this, keeping in mind that the stability of equilibria of the gLVE is determined by the leading eigenvalue of the `reduced interaction matrix'. This is the matrix of interactions restricted to the set of surviving species \cite{stone}, and is to be distinguished from the `original interaction matrix' among all species in the gLVE system.
		
		It was shown in \cite{baron_et_al_prl} that the eigenvalue spectrum of this reduced matrix typically consists of an elliptic bulk spectrum, plus, potentially, one isolated outlier eigenvalue. As also shown in \cite{baron_et_al_prl}, the onset of the linear instability coincides with the point in parameter space at which the bulk spectrum of the reduced matrix crosses the imaginary axis in the complex plane.
		
		We emphasise that the set of survivors is not a random subset of the initial pool. Instead, the survival of any particular species is determined by the original interaction matrix. As such, the set of survivors is correlated with the initial matrix, and as a consequence of the extinction dynamics, the interactions among survivors can have somewhat intricate statistics. This is explained in more detail in \cite{baron_et_al_prl} for the gLVE without generalised correlations in the original interactions. In particular, the reduced interaction matrix can display in-row, in-column and transpose row-column correlations, even if these are not present in the original interaction matrix.
		
		However, none of these correlations in the reduced interaction matrix affect the bulk eigenvalue spectrum. Instead the bulk eigenvalues of the reduced matrix are contained in the ellipse given by
		\be
		\frac{(1+ x)^2}{(1+\Gamma)^2} + \frac{y^2}{(1-\Gamma)^2 }<  \phi \sigma^2, \label{eq:ellipse}
		\ee
		where $x$ and $y$ are the real and imaginary parts of the eigenvalues respectively.
		
		The analysis of the reduced interaction matrix \cite{baron_et_al_prl} is for an original gLVE system without in-row, in-column and transpose row-column correlations. We have not extended this full analysis of the reduced matrix to systems in which there are generalised correlations already in the original system. However, it is reasonable to conjecture that the bulk spectrum of the reduced matrix would not be affected by in-row correlations in the original matrix, other than through a change of the fraction of surviving species. That is to say, we expect Eq.~(\ref{eq:ellipse}) to continue to hold, and all effects of in-row correlations in the original system to be contained in $\phi$. 
		
		The right-most point in the ellipse described by Eq.~(\ref{eq:ellipse}) the point $x=-1+\phi\sigma(1+\Gamma)$ and $y=0$. Therefore, the right edge of the ellipse crosses into the right half of the complex plane ($x=0$) when  $\phi\sigma(1+\Gamma)=1$. This then leads to Eq.~(\ref{eq:opper2}).
		
		\section{Cavity approach to dynamic mean-field theory}\label{app:cavity}
		In this appendix we present a cavity approach to deriving the effective process for the gLVE model with generalised correlations in the interaction matrix. Throughout this section we use an overbar to denote the disorder average. 
		
		Assume we have a solution $x_1(t), \dots, x_N(t)$ for the original Lotka--Volterra problem
		\be\label{eq:lv_cav}
		\dot x_i=x_i\left(1-x_i+\sum_{j (\neq i)= 1}^N\alpha_{ij}x_j+h_i(t)\right),
		\ee
		for given parameters $\mu, \gamma,\sigma^2$ and a fixed $N$. Now introduce an additional species $i=0$. This will change the dynamics for $x_1,\dots,x_N$, and we use $\tilde x_i$ for the trajectory of the system with species $0$ included [but we do not use a tilde for species $0$ itself, i.e., its abundance is $x_0(t)$]. For $i=1,\dots,N$ we then have
		\be\label{eq:lv_cav2}
		\frac{d}{dt}{\tilde x_i}(t)=\tilde x_i(t)\left(1-\tilde x_i(t)+\sum_{j=1}^N\alpha_{ij}\tilde x_j(t)+\alpha_{i0} x_0(t)+h_i(t)\right).
		\ee
		Writing $\alpha_{ij}=\frac{\mu}{N}+z_{ij}$ for $i,j=1,\dots,N$, we find
		\be\label{eq:lv_cav3}
		\frac{d}{dt}{\tilde x_i}(t)=\tilde x_i(t)\left[1-\tilde x_i(t)+\frac{\mu}{N}\sum_{j=1}^N \tilde x_j+\sum_{j=1}^Nz_{ij}\tilde x_j(t)+\alpha_{i0} x_0(t)+h_i(t)\right].
		\ee
		We identify (for $N\to \infty$)
		\be 
		\frac{1}{N}\sum_{j=1}^N \tilde x_j\rightarrow M(t),
		\ee
		and find
		\be\label{eq:lv_cav3}
		\frac{d}{dt}{\tilde x_i}(t)=\tilde x_i(t)\left[1-\tilde x_i(t)+\mu M(t)+\sum_{j=1}^Nz_{ij}\tilde x_j(t)+\alpha_{i0} x_0(t)+h_i(t)\right],
		\ee
		Without the term $\alpha_{i0}x_0(t)$ this would be just the standard dynamics, unperturbed by the presence of $x_0$. Relative to the system without species $0$ added this means that each species $i=1,\dots,N$ experiences a `perturbation' $\alpha_{i0} x_0(t)$ at time $t$. It is important to note that this happens to all species $i=1,\dots,N$ simultaneously, i.e., they are all perturbed.

		Treating these perturbations as small we can use linear response theory. We then have for $i=1,\dots,N$
		\be\label{eq:x_i_tilde}
		\tilde x_i(t)=x_i(t)+\sum_{j=1}^N \int^t dt' \frac{\delta x_i(t)}{\delta h_j(t')}\alpha_{j0}x_0(t').
		\ee
		The time evolution of $x_0$ on the other hand is given by
		\BE
		\frac{d}{dt} x_0(t) &=& x_0(t) \left[1-x_0(t)+\sum_{i=1}^N \alpha_{0i} \tilde x_i(t)+h_0(t)\right] \nonumber \\
		&=&  x_0(t) \left[1-x_0(t)+\mu M(t)+\sum_{i=1}^N z_{0i} \tilde x_i(t)+h_0(t)\right]
		\EE
		Using Eq.~(\ref{eq:x_i_tilde}) this is
		\BE
		\frac{d}{dt} x_0(t) &=& x_0(t) \left[1-x_0(t)+\mu M(t) +\underbrace{\sum_{i=1}^N z_{0i}  x_i(t)}_{\rm Term 1}+h_0(t)
		+\underbrace{\sum_{i=1}^N z_{0i}\sum_{j=1}^N \alpha_{j0}\int^t dt' \frac{\delta x_i(t)}{\delta h_j(t')}x_0(t')}_{\rm Term 2}\right].\label{eq:x0_dyn1}
		\EE
		
		\noindent \underline{Term 1:}\\
		We look at the term $\sum_i z_{0i}  x_i(t)$, noting that $z_{0i}$ has mean zero and variance $\sigma^2/N$. We also have $\overline{z_{0i}z_{0j}}=\frac{r}{N}\frac{\sigma^2}{N}$ for $i\neq j$ (in-row correlations).
		
		The variance and correlations in time of $\sum_i z_{0i} x_i(t)$ are,
		\BE
		\overline{\left(\sum_i z_{0i}  x_i(t)\right)\left(\sum_i z_{0i}  x_i(t')\right)} &=& \sum_{i,j=1}^N \overline{z_{0i}z_{0j}} x_i(t)x_j(t') \nonumber \\
		&=& \frac{\sigma^2}{N}\sum_{i=1}^N x_i(t)x_i(t')+\frac{r\sigma^2}{N^2}\sum_{i\neq j} x_i(t) x_j(t') \nonumber \\
		&=&\sigma^2 C(t,t')+\sigma^2 r M(t) M(t')+{\cal O}(N^{-1}).
		\EE
		Thus we can write Eq.~(\ref{eq:x0_dyn1}) as 
		\BE
		\frac{d}{dt} x_0(t) &=& x_0(t) \left[1-x_0(t)+\mu M(t) +\eta(t)+h_0(t)
		+\underbrace{\sum_{i=1}^N z_{0i}\sum_{j=1}^N \alpha_{j0}\int^t dt' \frac{\delta x_i(t)}{\delta h_j(t')}x_0(t')}_{\rm Term 2}\right],\label{eq:x0_dyn2}
		\EE
		with noise $\eta(t)$ of zero average and with
		\be
		\avg{\eta(t)\eta(t')}=\sigma^2 C(t,t')+\sigma^2 r M(t) M(t').
		\ee
		This reproduces the right noise term in the effective process as derived from the generating functional calculation [see Eq.~(\ref{eq:cfs})].
		
		\medskip
		
		\noindent \underline{Term 2}:\\
		We now deal with Term 2. Each contribution to the sum over $i$ and $j$ contains a factor $z_{0i}\alpha_{j0}=\frac{\mu}{N}z_{0i}+z_{0i}z_{j0}$. Given that $z_{0i}$ has average zero, we can discard the term $(\mu/N)z_{0i}$. Therefore, we are left with
		\be
		\mbox{Term 2}=\sum_{i,j=1}^N  z_{0i}z_{j0}\int^t dt' \frac{\delta x_i(t)}{\delta h_j(t')}x_0(t')
		\ee
		We distinguish between diagonal contributions ($i=j$), and non-diagonal terms ($i\neq j$). 
		\medskip
		
		\noindent{\em Diagonal terms.} We make the following replacement
		\be
		\sum_{i=1}^N \overline{z_{0i}z_{i,0}} \frac{\delta x_i(t)}{\delta h_i(t')}\to \Gamma\sigma^2 G(t,t')
		\ee
		This is justified because $\overline{z_{0i}z_{i0}}=\Gamma\frac{\sigma^2}{N}$ and $G(t,t')=\frac{1}{N}\sum_{i=1}^N  \frac{\delta x_i(t)}{\delta h_i(t')}$.
		\medskip
		
		\noindent{\em Non-diagonal term.}
		Now look at
		\be
		\sum_{i\neq j }^N \overline{z_{0i} z_{j0}}  \frac{\delta x_i(t)}{\delta h_j(t')}x_0(t').
		\ee
		We use 
		\be
		\overline{z_{0i}z_{j0}}=\frac{\gamma}{N}\frac{\sigma^2}{N}.
		\ee
		for $i\neq j$, to make the replacement
		\be\label{eq:non-diag}
		\sum_{i\neq j }^N \overline{z_{0i} z_{j0}}  \frac{\delta x_i(t)}{\delta h_j(t')}x_0(t')\to \frac{\gamma \sigma^2}{N^2}
		\sum_{i\neq j }^N G_{ij}(t,t')x_0(t'),
		\ee
		where we have introduced the shorthand $G_{ij}(t,t')=\frac{\delta x_i(t)}{\delta h_j(t')}$.
		
		The term $N^{-2}\sum_{i\neq j }^N G_{ij}(t,t')x_0(t')$ broadly describes the effect on $N^{-1}\sum_i x_i(t)$ if a perturbation $x_0(t')/N$ is simultaneously applied to \underline{all} species $j$ at the earlier time $t'$. However, we note that this is not the same as the term proportional to $\gamma$ that appears in the effective process in Eq.~(\ref{eq:eff_proc}). That term is given by $\sigma^2 \gamma \int_0^t dt' G(t,t') M(t')$.
		
		Let us try to understand why this is the case by returning to the generating functional in Eq.~(\ref{eqn:disorder_avg_OP}). From Eqs.~(\ref{eq:phi}) and (\ref{eqn: OP_1}), we see that the term proportional to $\gamma$ contributes a term to the action
		\begin{align}\label{eq:phigamma}
			\Phi_\gamma = i\gamma\sigma^2\int dt\int dt'  K(t',t)M(t)P(t') = - \frac{\gamma \sigma^2 }{N^3} \sum_{(i,j,k)} \int dt\int dt' x_i(t') \hat x_i(t) x_j(t) \hat x_k(t'),
		\end{align}
		where $(i,j,k)$ indicates the sum is over only combinations of $i,j,k$ in which the indices take pairwise distinct values.
		
		When we attempt to read off the effective process from the action of the generating functional, this term can be interpreted in two ways. Indeed, if we focus on the term in the sum in Eq.~(\ref{eq:phigamma}) for which $i = 0$, then we would collect the terms in the action contributing to the time evolution of $x_0$ as follows
		\begin{align}
			S_0 = i \int dt \hat x_0(t) \left[ \frac{\dot x_0(t)}{x_0(t)}  + \cdots +i \frac{\gamma \sigma^2 }{N^2} \sum_{(j,k)} \int dt'  x_j(t) \hat x_k(t') x_0(t')  \right],
		\end{align}
		where we write $\sum_i S_i = N (\Psi + \Phi + \Omega)$ [see Eq.~(\ref{eqn:disorder_avg_OP})], and this indeed yields the term in Eq.~(\ref{eq:non-diag}). 
		
		However, if we instead focus on the term in the sum in Eq,~(\ref{eq:phigamma}) where $k = 0$, and relabel the time coordinates such that $t \leftrightarrow t'$, then we obtain instead
		\begin{align}
			S_0 = i \int dt \, \hat x_0(t) \left[ \frac{\dot x_0(t)}{x_0(t)}  + \cdots + i\frac{\gamma \sigma^2 }{N^2} \sum_{(i,j)} \int dt'  x_i(t) \hat x_i(t') x_j(t')  \right],
		\end{align}
		which yields the term in Eq.~(\ref{eq:eff_proc}). We therefore see that there are two ways to interpret the term proportional to $\gamma$. Each of these terms would give the same behaviour for the ensemble of species. However, the term in Eq.~(\ref{eq:eff_proc}) is far more amenable to further analysis than that in Eq.~(\ref{eq:non-diag}), yet this former version is not readily accessible via the cavity method. This is therefore arguably an instance where the MSRJD formalism provides insight more readily than the cavity method.

	\end{appendix}

\end{document}